\documentclass{jetpl}
\usepackage {cite}
\twocolumn

\lat

\title{The Kohn-Luttinger effect and anomalous pairing in new superconducting systems and graphene}

\rtitle{The Kohn-Luttinger effect and anomalous pairing in new
superconducting systems and graphene}

\sodtitle{The Kohn-Luttinger effect and anomalous pairing in new
superconducting systems and graphene}

\author{M.\,Yu.\,Kagan$^{a,b}$,
V.\,V.\,Val'kov$^{c}$, V.\,A.
Mitskan$^{c,d}$, M.\,M.
Korovushkin$^{c}$}

\rauthor{M.\,Yu.\,Kagan V.\,V.\,Val'kov, V.\,A. Mitskan, M.\,M.
Korovushkin}

\sodauthor{}

\address{$^a$Kapitza Institute of Physical Problems, Russian Academy of Sciences, Moscow, 119334 Russia\\
$^b$Moscow State Institute of Electronics and Mathematics, National Research University Higher School of Economics, Moscow, 109028 Russia\\
$^c$Kirensky Institute of Physics, Siberian Branch, Russian Academy of Sciences, Krasnoyarsk, 660036 Russia\\
$^d$Reshetnev Siberian State Aerospace University, Krasnoyarsk, 660014 Russia\\
e-mail: kagan@kapitza.ras.ru; vvv@iph.krasn.ru }

\abstract{We present a review of theoretical investigations into
the Kohn-Luttinger nonphonon superconductivity mechanism in
various 3D and 2D repulsive electron systems described by the
Fermi-gas, Hubbard, and Shubin-Vonsovsky models. Phase diagrams of
the superconducting state are considered, including regions of
anomalous $s-$, $p-$, and $d-$wave pairing. The possibility of a
strong increase in the superconducting transition temperature
$T_c$ even for a low electron density is demonstrated by analyzing
the spin-polarized case or the two-band situation. The
Kohn-Luttinger theory explains or predicts superconductivity in
various materials such as heterostructures and semimetals,
superlattices and dichalcogenides, high-$T_c$ superconductors and
heavy-fermion systems, layered organic superconductors, and
ultracold Fermi gases in magnetic traps. This theory also
describes the anomalous electron transport and peculiar polaron
effects in the normal state of these systems. The theory can be
useful for explaining the origin of superconductivity and orbital
currents (chiral anomaly) in systems with the Dirac spectrum of
electrons, including superfluid $^3$He-A, doped graphene, and
topological superconductors.}

\begin{document}

\maketitle

\section{INTRODUCTION}

Conduction electrons in metals and positive ions form a
solid-state plasma that determines the complex of their electric,
galvanomagnetic, kinetic, and super-conducting properties. The
coupling between the subsystems of massive positive ions and light
fermions leads to the formation of electron-phonon interaction,
which determines the properties of the electron subsystem. In
particular, the effective interaction between electrons in the
solid-state plasma can differ substantially from the Coulomb
interaction of electrons in vacuum and can even change sign. This
important effect forms the basis of the electron-phonon mechanism
of the Cooper instability in traditional
superconductors~\cite{BCS57}.

The role of mediator, interaction with which initiates the
renormalization of the Coulomb interaction, can obviously be
played by any other subsystem. It is necessary only that the
interaction of the electron gas with such a subsystem leads to
polarization effects resulting in the production of electrons and
holes in the vicinity of the Fermi surface. In particular, in many
theoretical publications on high-temperature superconductors, the
role of such a mediator is played by collective excitations of the
subsystem of localized spins of copper ions. This effect is
associated with the spin-fluctuation mechanism of Cooper
instability leading to the formation of a superconducting phase
with $d$-wave type symmetry of the order parameter.

In the secondary quantization representation for fermions, the
operator of Coulomb interaction between electrons contains
nondiagonal terms initiating the polarization contributions to the
ground-state energy in higher orders of perturbation theory. These
contributions lead to renormalization of the Cooper interaction
between electrons. Therefore, the effective interaction of
electrons in such a substance can differ significantly from
electron-electron interaction in vacuum. This makes topical the
problem formulated for the first time by
Anderson~\cite{Anderson87}, associated with the possibility of
renormalization of the Coulomb interaction between electrons, so
that the effective electron-electron interaction becomes
attractive and not repulsive even when phonons are disregarded. In
other words, the problem involves searching for conditions in
which the above-mentioned polarization effects in the electron
plasma of solid metals reverse the sign of the resultant
interaction between electrons. The analytic solution to this
problem comes to calculating the effective pair interaction of
electrons with allowance for many-particle effects in the electron
ensemble. According to Anderson, an equally important problem is
explaining the peculiar properties of the normal state of many
strongly correlated electron systems above the superconducting
transition temperature, especially in the pseudogap state.

Considerable advances have been made in recent decades in
experimental and theoretical investigations of superconducting
systems with a nonphonon origin of Cooper pairing. The first
experimentally discovered systems with nontraditional triplet
$p$-wave pairing (the total spin of a Cooper pair is
$S_{\textrm{tot}} = 1$ and the orbital angular momentum is $l =
1$) were the superfluid $A$ and $B$ phases of $^3$He with low
superconducting transition temperatures
$T_c\sim1\,mK$~\cite{Vollhardt90,Volovik92,Volovik03}. Other
examples of systems in which $p$-wave pairing takes place are
$^6$Li$_2$ and $^{40}$K$_2$ molecules in magnetic taps under
Feshbach resonance conditions with ultralow superconducting
transition temperatures $T_c\sim10^{-7}$ to
$10^{-6}\,K$~\cite{Regal03,Schunck05}. It is assumed that
nontraditional $p$-wave pairing with superconducting transition
temperatures $T_c\sim0.5-1\,K$ takes place in some heavy-fermion
intermetallides such as U$_{1-x}$Th$_x$Be$_{13}$ and UNl$_2$Al$_3$
with high effective masses
$m^*\sim(100-200)m_e$~\cite{Ott84,Kromer98}. The $p$-wave paring
is often mentioned in connection with organic superconductors such
as $\alpha-$(BEDT-TTF)$_2$I$_3$ with
$T_c\sim5\,K$~\cite{Kuroki06}. Finally, $p$-wave pairing with
$T_c\sim1\,K$ is apparently achieved in ruthenates
Sr$_2$RuO$_4$~\cite{Maeno01} and probably in layered
dichalcogenides CuS$_2$--CuSe$_2$, semimetals, and semimetal
superlattices InAs-GaSb and PbTe-SnTe. Nontraditional
superconductors with singlet $d$-wave pairing
($S_{\textrm{tot}}=0,\,l=2$) include heavy-fermion intermetallide
UPt$_3$ with $m^*\sim200m_e$ and $T_c\sim0.5\,K$, as well as a
wide class of high-$T_c$ cuprate superconductors with
superconducting transition temperatures from $T_c=36\,K$ for
lanthanum-based compounds to $T_c=160\,K$ (the absolute
record-highest value of $T_c$ attained at present in cuprates) for
mercury-based superconductors obtained under pressure. Finally, we
should also mention (in connection with the nonphonon
superconductivity problem) new multiband superconductors such as
MgB$_2$~\cite{Nagamatsu01} and iron-arsenide-based superconductors
with the more traditional $s$-wave pairing, which were discovered
recently~\cite{Kamihara08}.

Apart from the problems of Cooper pairing in the above systems,
the still unsolved problems associated with the search for
superfluidity in 3D and, especially, 2D (thin films,
submonolayers) solutions of $^3$He in
$^4$He~\cite{Vollhardt90,Volovik92,Volovik03} and
superconductivity in doped graphene~\cite{Novoselov04} are of
considerable interest. Such systems are the most promising for
experimental and theoretical description of a wide class of
physical phenomena, including nontraditional superconductivity.

In particular, $^3$He submonolayers adsorbed on various substrates
such as a solid substrate (grafoil) or the free surface of
superfluid $^4$He permit the various correlation regimes in the
system (from ultrararefied Fermi gas to strongly correlated
fermion system~\cite{Kagan94_1}) to be achieved with variation of
the number density of particles in a wide range. This makes
solutions ideal objects for the development and testing of various
many-body methods in condensed matter theory.

Graphene is of considerable importance from the fundamental and
applied viewpoints due to its unique electronic
properties~\cite{Lozovik08,Kotov12}. In the vicinity of the Fermi
level, electrons in graphene exhibit linear dispersion, and the
energy gap between the valence band and the conduction band is
absent. For this reason, electrons can be described by the 2D
Dirac equation for zero-mass charged
quasiparticles~\cite{Wallace47}. The properties of such
quasiparticles (like reduced dimensionality, the spinor origin of
the spectrum, zero mass, and the absence of a gap in the spectrum)
lead to a number of nontrivial electronic effects that have no
analogs in other physical systems~\cite{Castro09}.

Such systems stimulated an intense search for alternative
superconducting pairing mechanisms, which are based on
correlations in the Fermi liquid. The Kohn-Luttinger
mechanism~\cite{Kohn65} proposed in 1965 is the most promising in
this respect. This mechanism presumes the transformation of
initial repulsive interaction of two particles in vacuum into
effective attraction in the presence of the fermionic background.
This review describes the main results on Kohn-Luttinger
superconductivity in repulsive Fermi systems, which have been
obtained in the last 50 years.

\section{ELECTRON GAS MODEL}

The Fermi gas model is the basic model for studying nonphonon
superconductivity mechanisms in low-density electron systems. In
the case of an attractive Fermi gas, the scattering length is
negative ($a<0$) and traditional $s$-wave pairing takes place
(total spin $S=0$ and orbital angular momentum $l=0$) with the
superconducting transition temperature
\begin{equation}\label{Tcs}
T^{s}_c\approx0.28\,\varepsilon_F\exp\biggl(-\frac{\pi}{2|a|p_F}\biggr).
\end{equation}
This result was obtained in~\cite{Gor'kov61} soon after the
formulation of the BCS theory~\cite{BCS57}. Result (\ref{Tcs})
differs from the classical BCS formula. Namely, the quantity
$0.28\,\varepsilon_F$ appears in the preexponential factor instead
of the Debye frequency $\omega_D$ typical of models of traditional
superconductors.

\begin{figure}[t]
\begin{center}
\includegraphics[width=0.48\textwidth]{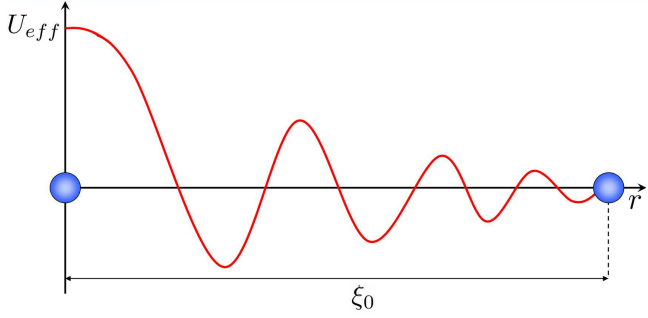}
\caption{Fig.~1. Friedel oscillations in the effective interaction of two particles due to polarization of the fermionic background: $\xi_0$ is the coherence length of a Cooper pair; $\xi_0\gg1/p_F$.} \label{friedel}
\end{center}
\end{figure}

In the model of a repulsive Fermi gas, the scattering length is
$a>0$ and superconductivity in the low-temperature range emerges
in accordance with the Kohn-Luttinger mechanism. The physical
reason for this pairing mechanism is associated with the effective
interaction of quasiparticles occurring as a result of
polarization of the fermionic background. In fact, due to the
sharp boundary existing in the momentum space, which is equal to
the diameter $2p_F$ of the Fermi sphere and separates the region
of filled states from the empty states, the effective interaction
of quasiparticles at the Fermi level does not decrease
exponentially, but has an alternating form (Friedel
oscillations~\cite{Freidel54}); in the 3D case, we have

\begin{equation}\label{oscillations}
U_{\textrm{eff}}(r)\sim\frac{\cos(2p_Fr)}{(2p_Fr)^3}.
\end{equation}
If the distance between two electrons in a Cooper pair is much
larger than the atomic spacing, effective interaction
(\ref{oscillations}) in the coordinate space has a large number of
maxima and minima (Fig.~\ref{friedel}). Then the integrated effect
determined by averaging of Friedel oscillations over such a
potential relief can generally result in the effective attraction
and the occurrence of superconductivity in the system.

\begin{figure*}[t]
\begin{center}
\includegraphics[width=0.98\textwidth]{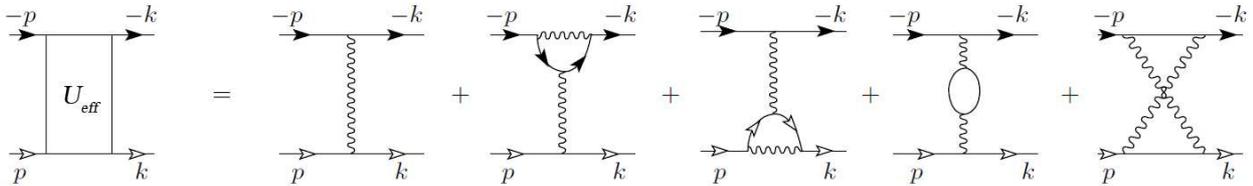}
\caption{Fig.~2. First- and second-order diagrams for effective
interaction $U_{\textrm{eff}}$. Solid lines with light (dark)
arrows correspond to the electron Green's function with a spin
projection of $+\frac12 ~(-\frac12)$. Wavy lines reflect the
initial interaction.} \label{diagrams_2order}
\end{center}
\end{figure*}

Kohn and Luttinger~\cite{Kohn65}, who investigated a 3D electron
gas, were the first to pay attention to this superconductivity
mechanism. They showed that the effective interaction in the first
two orders of perturbation theory can be described by the sum of
five diagrams shown in Fig.~\ref{diagrams_2order}. The first
diagram corresponds to the initial interaction of two electrons in
the Cooper channel. The next four (Kohn-Luttinger) diagrams
reflect second-order processes and take into account the
polarization effects in the filled Fermi sphere. In the case of a
low-density Fermi gas and a short-range potential, the
contribution to the effective interaction is determined only by
the fourth (exchange) diagram; in the first two orders of
perturbation theory in gas parameter $ap_F$, the expression for
$U_{\textrm{eff}}$ can be written in the form
\begin{equation}
U_{\textrm{eff}}(\textbf{p},\textbf{k})=ap_F+(ap_F)^2\Pi(\textbf{p}+\textbf{k}),
\end{equation}
where $\Pi(\textbf{p}+\textbf{k})$ is the static polarization
operator described by the standard Lindhard
function~\cite{Lindhard54,Ashcroft79}
\begin{equation}\label{Lindhard}
\Pi(\textbf{p}+\textbf{k})=\frac1N\sum_{\textbf{p}_1}\frac{n_F(\varepsilon_{\textbf{p}_1-\textbf{p}-\textbf{k}})-
n_F(\varepsilon_{\textbf{p}_1})} {\varepsilon_{\textbf{p}_1}-
\varepsilon_{\textbf{p}_1-\textbf{p}-\textbf{k}}},
\end{equation}
which is responsible for charge screening in the case of the electron gas in a metal. Here, $\varepsilon_{\textbf{p}}=\displaystyle\textbf{p}^2/{2m}$ is the energy spectrum,
\[ n_F(x)=\biggl(\exp(\frac{x-\mu}{T})+1\biggr)^{-1} \]
is the Fermi-Dirac distribution function, and $\mu$ is the chemical potential.

It was noted in early publications by Migdal~\cite{Migdal58} and
Kohn~\cite{Kohn59}, as well as in~\cite{Kohn65}, that at low
temperatures ($T\ll\varepsilon_F$), the polarization operator
contains, in addition to the regular part, a singular part also,
which is known as the Kohn singularity and has the following form
in the 3D space:
\begin{equation}
\Pi_{\textrm{sing}}(\widetilde{q})\sim(\widetilde{q}-2p_F)\ln|\widetilde{q}-2p_F|,
\end{equation}
where $\widetilde{q}=|\textbf{p}+\textbf{k}|$ for the fourth
exchange diagram in Fig.~\ref{diagrams_2order}. As we pass to the
coordinate space, singular part $\Pi_{\textrm{sing}}$ leads to
Friedel oscillations (\ref{oscillations}) in the effective
interaction (see Fig.~\ref{friedel}). Thus, the purely repulsive
short-range potential acting between two particles in vacuum
induces the effective interaction in the electron gas with the
competition between repulsion and attraction. It turns out that
the singular part in $U_{\textrm{eff}}$ operates in favor of
attraction, ensuring a contribution that always exceeds the
repulsive contribution associated with the regular part of
$U_{\textrm{eff}}$ for the orbital angular momenta $l\neq0$ of a
pair. This leads to superconducting instability with the critical
temperature $T_{c,l}\sim\exp(-l^4)$ for large orbital angular
momenta $l\gg1$. In this case, conventional pairing in the
$s$-wave channel ($l=0$) is suppressed by purely Coulomb repulsion
associated with the main maximum in $U_{\textrm{eff}}$ (see
Fig.~\ref{friedel}), and superconductivity is observed for large
values of orbital angular momentum $l\gg1$. It should be noted
that for $l\neq0$, the role of the main maximum is suppressed by
the centrifugal potential, which improves the conditions for the
occurrence of superconductivity.

Thus, publication~\cite{Kohn65} led to the nontrivial conclusion
that Fermi systems do not exist in the normal state at zero
temperature because any 3D electron system with the initial
repulsive interaction between particles at very low temperatures
is unstable to the transition to the superconducting state with a
large orbital angular momentum ($l\gg1$) of the relative motion of
a Cooper pair. However, the estimates for the superconducting
transition temperature obtained in~\cite{Kohn65} for the realistic
parameters of electron systems in a metal and for superfluid
$^3$He gave very low values of the superconducting transition
temperature ($T_{c,d}\sim10^{-16} K$ for $^3$He and
$T_{c,d}\sim10^{-11} K$ for the metal plasma for the value of
$l=2$ considered in~\cite{Kohn65}). Such a low $T_c$ value was one
of the reasons the Kohn-Luttinger mechanism was overlooked by
researchers for a long time.

It was shown in later publications~\cite{Fay68,Kagan88} that the
superconducting transition temperature was underestimated
in~\cite{Kohn65} because the asymptotic expression for large
orbital angular momenta $l\gg1$ was used. Indeed, for $l=1$, exact
analytic calculations show that the contributions to
$U_{\textrm{eff}}$ corresponding to attraction of quasi-particles
again prevail over the repulsive contributions. As a result, a 3D
repulsive Fermi gas turns out to be unstable to the
superconducting transition with triplet $p$-wave pairing at the
superconducting transition
temperature~\cite{Fay68,Kagan88,Baranov92b,Baranov96}, which is
determined by the principal exponential:
\begin{equation}\label{Tcp}
T_c\approx\varepsilon_F\exp\biggl(-\frac{5\pi^2}{4(2\ln2-1)(ap_F)^2}\biggr)
=\varepsilon_F\exp\biggl(-\frac{13}{\lambda^2} \biggr),
\end{equation}
where $\lambda=\displaystyle{2ap_F}/{\pi}$ is the effective 3D
Galitskii gas parameter~\cite{Galitskii58}. It should be noted
that for $l=1$, the contribution of the Kohn singularity only
increases $T_{c,p}$, but does not play a decisive role in the
occurrence of superconductivity.

It was shown in~\cite{Kagan89} that the superconducting transition
temperature can appreciably be elevated even for low electron
densities by placing a system of neutral particles into a magnetic
field. This is due to the fact that in contrast to $s$-wave
pairing, paramagnetic suppression of superconductivity does not
take place in the $p$-wave channel and the value of $T_{c,p}$ may
increase due to the enhancement of the effective interaction and
due to the specific form of the Kohn singularity. In this case,
the critical temperature $T_{c,p}$ correspond to the so-called
$A_1$ phase, in which a Cooper pair is formed by two "up" spins,
while the effective interaction for them is prepared by two "down"
spins.

In the case of a low-density 2D repulsive Fermi gas, the effective
interaction in the first two orders of perturbation theory in the
gas parameter has the form~\cite{Chubukov93,Efremov00b}
\begin{equation}
U_{\textrm{eff}}(\textbf{p},\textbf{k})=f_0+f_0^2\Pi(\textbf{p}+\textbf{k}),
\end{equation}
where $f_0=\displaystyle{1}/{2\ln(p_Fr_0)}$ is the 2D Bloom gas
parameter~\cite{Bloom75}, $\Pi(\textbf{p}+\textbf{k})$ is the 2D
polarization operator, and $r_0$ is the range of the potential.

In the 2D situation, the effective interaction in the coordinate
space also contains Friedel oscillations

\begin{equation}\label{oscillations2D}
U_{\textrm{eff}}(r)\sim f_0^2\frac{\cos(2p_Fr)}{(2p_Fr)^2},
\end{equation}
which are much stronger than oscillations (\ref{oscillations}) in
the 3D case. However, the 2D Kohn singularity in the momentum
space has one-sided character~\cite{Afanasiev62}:
\begin{equation}
\Pi_{\textrm{sing}}(\widetilde{q})\sim
f_0^2\textrm{Re}\sqrt{\widetilde{q}-2p_F}=0
\end{equation}
for $\widetilde{q}=({\bf p}+{\bf k})\leq2p_F$ and is ineffective
for the problem of superconductivity. Thus, a 2D repulsive Fermi
gas remains in the normal state at least in the first two orders
of perturbation theory in gas parameter $f_0$. Nevertheless, it
was shown in~\cite{Chubukov93} that superconducting $p$-wave
pairing appears in the next (third) order of perturbation theory
in $f_0$, in which the singular contribution to the effective
interaction has the form

\begin{equation}
\Pi_{\textrm{sing}}(\widetilde{q})\sim
f_0^3\textrm{Re}\sqrt{2p_F-\widetilde{q}}.
\end{equation}
Exact calculations~\cite{Efremov00a} of the superconducting
transition temperature taking into account all irreducible
third-order diagrams leads to the expression

\begin{equation}\label{Tcp3order}
T_c\sim\varepsilon_F\biggl(-\frac{1}{6.1f_0^3} \biggr).
\end{equation}
In this case, the superconducting transition temperature is
estimated as $10^{-4}$K~\cite{Chubukov93,Efremov00a} for the
limiting densities for which the Fermi-gas description is still
applicable. This estimate is closer to the realistic values
predicted for $^3$He monolayers on the $^4$He surface~\cite{Oh94}.

Another possibility to sharply increase $T_c$ at low density is
associated with the analysis of the two-band situation or a
multilayer system. In this case, the role of "up" spins is played
by electrons from the first band (layer), while the role of "down"
spins is played by electrons of the second band (layer). The
coupling between electrons of the two bands is accomplished via
interband Coulomb interaction. As a result, the following
mechanism of superconducting pairing is possible: electrons of one
species form a Cooper pair by polarizing electrons of the other
species~\cite{Kagan91,KaganValkov11}. This mechanism of
interaction is also effective in quasitwo-dimensional systems.

It should be noted that some authors~\cite{Woelfle11,Baranov93}
also studied the effect of split-off energy bands on the
properties of the normal state in the basic models for repulsive
Fermi systems. For example, nontrivial corrections to the
Galitskii-Bloom Fermi-gas expansion appear due to antibound
states~\cite{Woelfle11} in the 2D Hubbard model or due to a
singularity in the Landau quasiparticle $f$-function in a
repulsive 2D Fermi gas at low electron density~\cite{Baranov93}.
It was shown in~\cite{Woelfle11,Baranov93}, however, that these
corrections do not destroy the Landau Fermi-liquid picture in the
3D or in the 2D case.

\section{HUBBARD MODEL}

The Hubbard model~\cite{Hubbard63}, which is one of fundamental
models for describing peculiar properties of cuprates, has become
very popular in connection with the discovery of high-temperature
superconductivity~\cite{Bednorz86}. The Hubbard model is a special
case of a more general model of interacting electrons whose band
structure can be described using the tight-binding method; in
fact, the Hubbard model is the minimal model taking into account
the band motion of electrons in a solid along with strong
electron-electron
interaction~\cite{Izyumov94,Izyumov95,Georges96,Tasaki98,VVVSGO01}.
This model is especially important for describing narrow-band
metals~\cite{Efremov00a}. In the secondary quantization
representation, the Hamiltonian of such a model can be written in
the form

\begin{eqnarray}\label{HubbardHamiltonian}
\hat{H}&=&\sum\limits_{f\sigma}(\varepsilon-\mu)
n_{f\sigma}+\sum\limits_{fm\sigma}t_{fm}c_{f\sigma}^\dag
c_{m\sigma} +U\sum\limits_f
n_{f\uparrow}n_{f\downarrow},\nonumber\\
\end{eqnarray}
where $c_{f\sigma}^\dag (c_{f\sigma})$ is the creation
(annihilation) operator for an electron with spin projection
$\sigma=\pm1/2$ at the $f$ site, $\varepsilon$ is the single-site
electron energy, and  $\mu$ is the chemical potential of the
system. In expression (\ref{HubbardHamiltonian}),
\[n_f=\sum\limits_\sigma n_{f\sigma}=\sum\limits_\sigma c_{f\sigma}^\dag c_{f\sigma}\]
is the operator of the particles density at site $f$, matrix
element $t_{fm}$ determines the intensity of electron hoppings
from site $f$ to site $m$, and $U$ is the Coulomb interaction
parameter for two electrons located at the same site and having
opposite projections of the spin moment (Hubbard repulsion).

\begin{figure}[t]
\begin{center}
\includegraphics[width=0.48\textwidth]{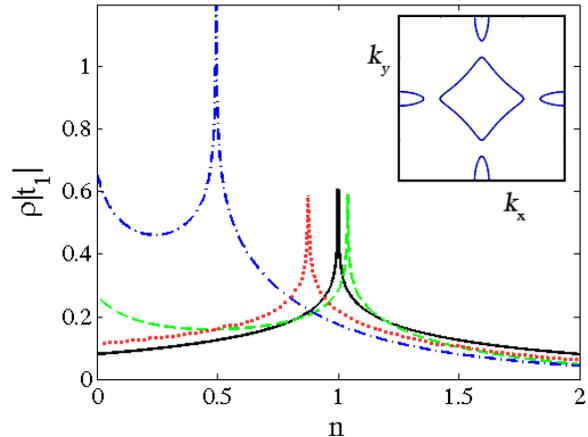}
\caption{Fig.~3. Modification of the electron density of states
and the shift of the Van Hove singularity in the Hubbard model on
a square lattice upon a change in the hopping integral:
$t_2=t_3=0$ (solid curve), $t_2=0.15,~t_3=0$ (dotted curve),
$t_2=0.15,~t_3=0.1$ (dashed curve), and $t_2=0.44,~t_3=0$
(dot-and-dash curve). The inset shows the formation of a
multisheet Fermi contour for the set of parameters
$t_2=0.44,~t_3=-0.1$, and $\mu=2$ (all parameters are given in
units of $|t_1|$).} \label{DOS_Hubbard}
\end{center}
\end{figure}

Since a large body of experimental data indicated that the main
dynamics of Fermi excitations in cuprates evolves in the CuO$_2$
planes, the 2D Hubbard model on a simple square lattice was mainly
used to describe the nonphonon mechanisms of high-temperature
superconductivity. In the momentum space, the Hamiltonian of the
Hubbard model has the form
\begin{eqnarray}\label{Hubbard_momentum}
\hat{H}
&=&\sum\limits_{\textbf{p}\sigma}(\varepsilon_{\textbf{p}}-\mu)
c^{\dagger}_{\textbf{p}\sigma}c_{\textbf{p}\sigma} +
U\sum_{\textbf{p}\textbf{p'}\textbf{q}}c^{\dagger}_{\textbf{p}\uparrow}
c^{\dagger}_{\textbf{p'}+\textbf{q}{\downarrow}}c_{\textbf{p}+\textbf{q}{\downarrow}}
c_{\textbf{p'}{\uparrow}},\nonumber\\
\end{eqnarray}
where the electron energy taking into account distant hoppings,
whose intensity is determined by parameters $t_2$ and $t_3$, is
described by the expression
\begin{eqnarray}
\varepsilon_{\textbf{p}}&=&2t_1(\textrm{cos}\,p_xa+\textrm{cos}\,p_ya)+
4t_2\textrm{cos}\,p_xa\,\textrm{cos}\,p_ya +\nonumber\\
&+&2t_3(\textrm{cos}\,2p_xa+\textrm{cos}\,2p_ya),\label{Hubbard_spectra}
\end{eqnarray}
where $d$ is the intersite distance. It should be noted that in
simulating electron spectrum (\ref{Hubbard_spectra}) and
constructing the phase diagram of the superconducting state in the
Hubbard model, it becomes important to exceed the bounds of the
nearest-neighbor approximation ($t_2\neq0,\,t_3\neq0$). This is
due to the fact that the main contribution to the effective
coupling constant comes from the interaction of electrons on the
Fermi surface with a geometry depending on the structure of the
energy spectrum. The fact that the inclusion of distant hoppings
shifts the Van Hove singularity in the density of electron states
from half-filling ($n=1$) to the regions of lower or higher
electron densities (Fig.~\ref{DOS_Hubbard}) also plays an
important role. It should be noted that the inclusion of hoppings
to the third coordination sphere of the square lattice
($t_3\neq0$) can lead to a qualitative change in the Fermi surface
geometry (in particular, to the formation of a multisheet Fermi
contour; see the inset to Fig.~\ref{DOS_Hubbard}).

Thus, the allowance for distant hoppings can modify the phase
diagram determining the range of super-conducting states with
various types of the order parameter symmetry.

In the Hubbard model, perturbation theory can be constructed in
two limiting cases, viz., the Born approximation with a shallow
potential well ($U\ll W$, where $W=2zt$ is the bandwidth and $z$
is the number of the nearest neighbors) and an arbitrary electron
density, and in the case of strong coupling ($U\gg W$) and a low
electron density. The application of the weak coupling
approximation ($U\ll W$) for analyzing the possibility of
Kohn-Luttinger pairing makes it possible to calculate
$U_{\textrm{eff}}$ for the Cooper channel in the electron density
range $n\sim1$ (in the vicinity of half-filling) using diagrams of
no higher than the second order in the interaction (see
Fig.~\ref{diagrams_2order}). In the opposite limit of strong
coupling ($U\gg W$), the use of diagrams of only the first and
second orders is valid only for a low electron density $n<<1$, for
which the Galitskii-Bloom Fermi gas expansion is
valid~\cite{Galitskii58,Bloom75}.

In~\cite{Baranov92a}, the authors analyzed the conditions of the
Kohn-Luttinger superconductivity realization in the 2D Hubbard
model with Hamiltonian (\ref{Hubbard_momentum}) in the
weak-coupling limit ($U\ll W$) in the nearest neighbor
approximation ($t_2=t_3=0$) at the low electron densities
($p_Fd\ll1$):
\begin{eqnarray}
\varepsilon_{\textbf{p}}-\mu&=&2t_1(\textrm{cos}\,p_xd+\textrm{cos}\,p_yd)-\mu\approx\nonumber\\
&\approx&\frac{p^2-p^2_F}{2m}-\frac{(p_x^4+p_y^4)d^2}{24m}+\frac{(p_x^6+p_y^6)d^4}{720m},
\end{eqnarray}
where $m = 1/(2t_1d^2)$ is the band mass of an electron. It can be
seen that the initial electron spectrum in the chosen
approximation for $p_Fd\ll1$ almost coincides with the spectrum of
a noninteracting Fermi gas, and the Hubbard Hamiltonian itself is
identically equivalent to the Hamiltonian of a weakly nonideal
Fermi gas with a short-range repulsion between particles. To
verify the possibility of a superconducting transition in this
approximation, the effective initial vertex for the Cooper channel
was calculated up to the second order of perturbation theory
inclusively:
\begin{eqnarray}\label{Gamma_waveU}
&&U_{\textrm{eff}}(\textbf{p},\textbf{k})=U+U^2\Pi(\textbf{p}+\textbf{k}),
\end{eqnarray}
where $\Pi(\textbf{p}+\textbf{k})$ is polarization operator
(\ref{Lindhard}).

To solve the Bethe-Saltpeter integral equation,
in~\cite{Baranov92a} the eigenfunctions of the irreducible
representations of symmetry group $C_{4v}$ of the square lattice
have been used. This group is known to have five irreducible
representations~\cite{Landau89}, for each of which the integral
equation has its own solution. Among these representations, there
are four 1D representations $A_1,\,A_2,\,B_1$, and $B_2$, which
correspond to singlet pairing, as well as a 2D representation $E$
corresponding to triplet pairing. The explicit form of orthonormal
functions $g_{m}^{(\alpha)}(\phi)$ (superscript "$\alpha$" denotes
an irreducible representation, $m$ is the number of the basis
function of the representation $\alpha$, and $\phi$ is the angle
characterizing the direction of the momentum $\hat{\textbf{p}}$
lying on the Fermi contour relative to the $p_x$ axis) and the
symmetry classification of superconducting order parameter
$\Delta^{(\alpha)}(\phi)$ are defined as
\begin{eqnarray}\label{harmon}
&&A_1\rightarrow~g_{m}^{(s)}(\phi)=\frac{1}{\sqrt{(1+\delta_{m0})\pi}}\,
\textrm{cos}\,4m\phi,~~m\in[\,0,\infty),\label{invariants_s}\nonumber\\
&&A_2\rightarrow~g_{m}^{(s_{ext})}(\phi)=\frac{1}{\sqrt{\pi}}\,\textrm{sin}\,
4(m+1)\phi,\label{invariants_s1}\nonumber\\
&&B_1\rightarrow~g_{m}^{(d_{xy})}(\phi)=\frac{1}{\sqrt{\pi}}\,
\textrm{sin}\,(4m+2)\phi,\label{invariants_dxy}\\
&&B_2\rightarrow~g_{m}^{(d_{x^2-y^2})}(\phi)=\frac{1}{\sqrt{\pi}}\,
\textrm{cos}\,(4m+2)\phi,\label{invariants_dx2y2}\nonumber\\
&&E~\rightarrow~g_{m}^{(p)}(\phi)=\frac{1}{\sqrt{\pi}}\,(A\,\textrm{sin}\,
(2m+1)\phi+\nonumber\\
&&\qquad\qquad\qquad\qquad+B\,\textrm{cos}\,(2m+1)\phi)\label{invariants_p}\nonumber.
\end{eqnarray}
To solve the problem of superconducting pairing, function
$U_{\textrm{eff}}(\textbf{p}, \textbf{k})$ was expanded into a
series with functions (\ref{harmon}), after that the sign of
expressions for $U_{\textrm{eff}}^{\alpha}$  was analyzed. As a
result, it was shown that the 2D electron system described by the
Hubbard model for a small occupancy and for $U\ll W$ is unstable
to the pairing with the $d_{xy}-$wave type of symmetry of the
order parameter $\Delta(\phi)$.

\begin{figure}[t]
\begin{center}
\includegraphics[width=0.4\textwidth]{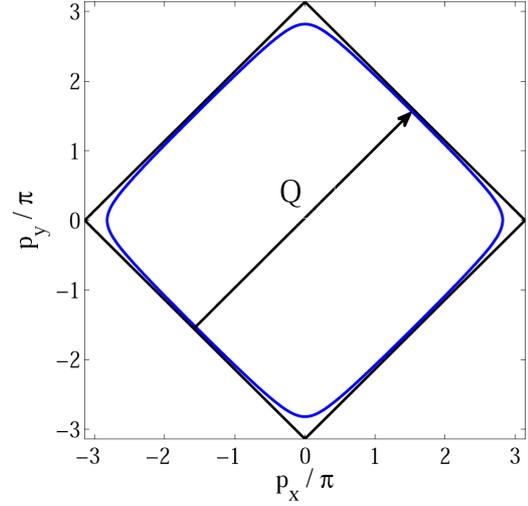}
\caption{Fig.~4. Fermi surface in the case of a nearly half-filled
band ($n\approx1$) in the 2D Hubbard model on a square lattice,
where $\textbf{Q}=(\pi/d,\pi/d)$ is the nesting vector.}
\label{nesting}
\end{center}
\end{figure}

The weak-coupling limit ($U<W$) in the 3D and 2D Hubbard models in
the vicinity of half-filling was analyzed
in~\cite{Scalapino86,Kozlov89}. In the 2D case~\cite{Kozlov89},
for $n\approx1$ in the nearest neighbor approximation, the
electron spectrum becomes
quasi-hyperbolic~\cite{Dzyaloshinskii88},
\begin{equation}\label{epsilon_p}
\varepsilon_{\textbf{p}}\approx\pm\frac{p_x^2-p_y^2}{2m}
\end{equation}
for small values of $p_xd < 1$ and $p_yd < 1$ in the vicinity of
corner points $(0,\pm \pi)$ and $(\pm \pi,0)$, at which the Fermi
surface almost touches the Brillouin zone (Fig.~\ref{nesting}). In
expression (\ref{epsilon_p}), the band mass is $m = 1/(2t_1d^2)$
as before. It is well known that the density of electron states in
these regions near the Van Hove singularity is logarithmically
large ($g(E)\displaystyle\sim\ln(t/|\mu|)$~), where $\mu\ll t$ is
the chemical potential in the vicinity of half-filling). It can be
seen from Fig.~\ref{nesting} that there are two almost planar
regions near the Fermi surface, which satisfy the ideal nesting
condition for the exactly half-filled state ($n=1$):
\begin{equation}
\varepsilon_{\textbf{p}+\textbf{Q}}=-\varepsilon_{\textbf{p}},
\end{equation}
where $\textbf{Q}=(\pi/d,\pi/d)$ is the nesting vector for a 2D
square lattice. In these regions, the polarization operator is
enhanced by the Kohn singularity, as well as by the Van Hove
singularity, and has the form~\cite{Kozlov89,Dzyaloshinskii88}
$\Pi(\textbf{Q})\displaystyle\sim\ln^2(t/|\mu|)$. In this case,
the parameter of perturbation theory is the quantity
\begin{equation}
f_0=\frac{U}{8\pi t}\ll1,
\end{equation}
and the expression for the effective interaction in the second
order of perturbation theory in $f_0$ has the form~\cite{Kozlov89}
\begin{equation}
U_{\textrm{eff}}\sim f_0+f_0^2\ln^2\frac{t}{|\mu|}.
\end{equation}
Since the expression for the Cooper loop for $n\approx1$ contains,
apart from the conventional Cooper logarithm, the logarithm of the
Van Hove singularity as well, we can ultimately write
\begin{eqnarray}
L(\xi_{\textbf{p}})=\displaystyle
\frac1N\sum_{\textbf{p}}\frac{\tanh(\xi_{\textbf{p}}/2T)}{2\xi_{\textbf{p}}}\sim\ln\frac{\mu}{T}
\ln\frac{t}{|\mu|},
\end{eqnarray}
where $\xi_{\textbf{p}}=\varepsilon_{\textbf{p}}-\mu$. Therefore,
the equation for the temperature of the superconducting transition
to the phase with the $d_{x^2-y^2}-$wave symmetry of the order
parameter derived in~\cite{Kozlov89} in the main logarithmic
approximation has the form
\begin{equation}
f_0^2\ln^3\frac{t}{|\mu|}\ln\frac{\mu}{T_c}\sim1.
\end{equation}
Hence, the superconducting transition temperature is given by
\begin{equation}\label{Tcdx2y2}
T^{d_{x^2-y^2}}_c\sim\mu\exp\Biggl(-\frac{1}{f_0^2\ln^3(t/|\mu|)}\Biggr).
\end{equation}
This expression shows that the small value of $f_0^2$ for
$f_0\,\ll\,1$ is compensated by the large value of
$\displaystyle\ln^3\frac{t}{|\mu|}\gg1$.

The results obtained in~\cite{Baranov92a} on $d_{xy}-$wave pairing
for $n\lesssim0.6$ and $d_{x^2-y^2}-$wave pairing for
$n\sim1$~\cite{Scalapino86,Kozlov89} in the strong coupling limit
were subsequently confirmed by other authors too.
In~\cite{Hlubina99}, the phase diagram of the superconducting
state was constructed in the 2D Hubbard model at low and
intermediate electron densities; this diagram reflected the
dependence of the competition of various symmetry types of the
order parameter on integral $t_2$ of electron hopping to the
next-to-nearest neighbors sites. The phase diagram obtained in the
second order of perturbation theory shows that for $t_2=0$, in the
range of low electron densities $0<n<0.52$, superconductivity with
the $d_{xy}-$wave type of the order parameter symmetry is realized
in the first two orders of perturbation theory; in the interval
$0.52<n<0.58$, the ground state corresponds to a phase with
$p-$wave pairing, while for $n>0.58$, $d_{x^2-y^2}-$wave pairing
appears. Analogous results were obtained in~\cite{Zanchi96} using
the renormalized group method.

In the immediate vicinity of the half-filling ($0.95<n<1$), where
strong competition between superconductivity and
antiferromagnetism takes place, the problem of the Cooper
instability was considered
in~\cite{Dzyaloshinskii88,Zheleznyak97}. In these publications,
the so-called parquet diagrams were summed up and the following
relation was obtained for $\mu\sim T_c$ :
\begin{equation}
f_0^2\ln^4\frac{t}{|\mu|}\sim f_0^2\ln^4\frac{t}{T_c}\sim1.
\end{equation}
This relation leads to an elegant estimate of the maximal superconducting transition temperature:
\begin{equation}\label{Tcdx2y2_parquet}
T^{d_{x^2-y^2}}_c\sim
t\exp\Biggl(-\frac{\textrm{const}}{\sqrt{f_0}}\Biggr).
\end{equation}

It should be noted that the maximal superconducting transition
temperature in the 2D Hubbard model was also obtained
in~\cite{Raghu10} in the regime $U/W\sim1$ for optimal electron
concentrations $n\sim0.8-0.9$. According to the estimate obtained
in~\cite{Raghu10}, the superconducting transition temperature at
the maximum can reach desirable values
$T^{d_{x^2-y^2}}_c\approx100\,K$, which are realistic for
optimally doped cuprate superconductors.

\section{SHUBIN-VONSOVSKY MODEL }

The important question concerning the role of full Coulomb
interaction in nonphonon superconductivity mechanisms, which in
fact includes not only short-range Hubbard repulsion, but also the
long-range component, was considered in~\cite{Alexandrov11}. The
authors of~\cite{Alexandrov11} considered the 3D jelly model for
realistic values of electron densities with $r_S\leq20$, where
\begin{equation}
r_S=\frac{1.92}{p_Fa_B}
\end{equation}
is the Wigner-Seitz correlation radius and
$a_B=\displaystyle{\varepsilon_0}/{me^2}$ is the Bohr radius of
electron ($\hbar=1$). In calculation of the effective interaction,
the contributions of the first and second orders of perturbation
theory associated with all diagrams in Fig.~\ref{diagrams_2order}
were taken into account. It was noted in~\cite{Alexandrov11} that
previous investigations of Kohn-Luttinger superconductivity were
limited to the inclusion of only short-range Coulomb interaction
$U$ of electrons in view of computational difficulties associated
with taking into account the Fourier transform of the long-range
Coulomb repulsion $V_{\textbf{q}}$, which depends on
${\textbf{q}}$ in the first- and second-order diagrams. As a
result, the strong initial Coulomb repulsion in the first order of
perturbation theory (first diagram in Fig.~\ref{diagrams_2order})
was disregarded, and the contribution to the effective interaction
of electrons in the Cooper channel was due only to the last
exchange diagram in Fig.~\ref{diagrams_2order}. This contribution
was attractive by nature and ensured $p-$wave pairing in the 3D
case~\cite{Fay68,Kagan88} and the $d-$wave pairing in the 2D case
~\cite{Baranov92b,Raghu10} in the Hubbard model.

In~\cite{Alexandrov11}, the long-range Coulomb interaction
$V_{\textbf{q}}$ was chosen in the form of the Fourier transform
of the Yukawa potential
\[V(r)=\displaystyle\frac{e^2}{r}\exp(-\kappa r),\] $V_{\textbf{q}}$
has the following standard form in the 3D case:
\begin{equation}\label{screening}
V_{\textbf{q}}=\frac{4\pi e^2}{q^2+\kappa^2},
\end{equation}
where $\kappa$ is the inverse screening length. It was concluded
in~\cite{Alexandrov11} from the results of calculations that small
and intermediate values of Hubbard repulsion $U$ in the presence
of the long-range part of Coulomb interaction (\ref{screening}) do
not induce realization of the Cooper instability in 3D and 2D
Fermi systems in the $p-$wave and $d-$wave channels, irrespective
of the value of the small screening length. The pairing that
occurs for large orbital angular momenta ($l\geq3$) leads to
almost zero values of the superconducting transition temperature
for any realistic value of the Fermi energy. Thus, anomalous
pairing associated with strong Coulomb repulsion is impossible
according to the authors of~\cite{Alexandrov11}, because the
corresponding energy of condensation for Cooper pairs is several
times lower than the energy of condensation associated with
electron-phonon interaction.

The rising interest in the role of the long-range inter-site
Coulomb correlations in the structure of the phase diagram of
high-$T_c$ superconductors has made the extended Hubbard model
popular. This model takes into account not only one-site Hubbard
repulsion, but the interaction of electrons at different sites of
the crystal lattice (in the Russian literature, this model is
usually referred to as the Shubin-Vonsovsky
model~\cite{Shubin34}).

\begin{figure}[t]
\begin{center}
\includegraphics[width=0.43\textwidth]{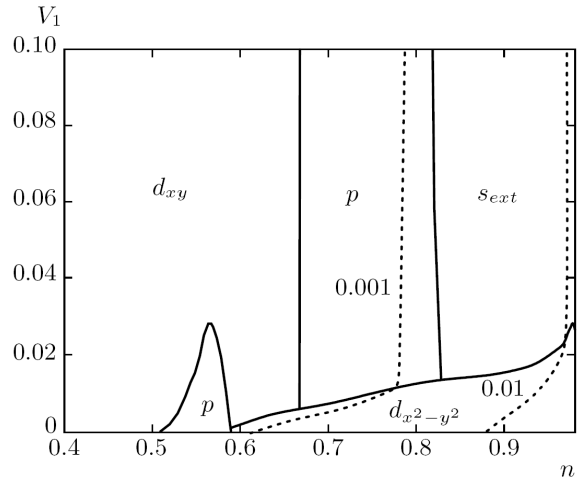}
\caption{Fig.~5. Phase diagram in the Shubin-Vonsovsky model for
$t_2=t_3=0,~U=|t_1|$ and $V_2/V_1=0$. The intersite Coulomb
interaction is taken into account only in the first order of
perturbation theory. For all points belonging to the same dotted
curve, the value of $\lambda$ is constant and marked by the
corresponding numeral.} \label{PD1}
\end{center}
\end{figure}

Historically, this model was formulated almost immediately after
the origination of quantum mechanics and is a predecessor of some
important models in condensed matter theory (in particular, the
$s-d(f)-$model and the Hubbard model). The Shubin-Vonsovsky model
was actively used in analyzing polaron states in
solids~\cite{Vonsovsky79} and the metal-insulator
transition~\cite{Zaitsev80}, as well as in studying the effect of
the intersite Coulomb interaction on the superconducting
properties of strongly correlated
systems~\cite{Zaitsev88,Zaitsev04,Valkov11}.

In the Wannier representation, the Hamiltonian of the Shubin-Vonsovsky model can be written in the form
\begin{eqnarray}\label{SVHamiltonian}
\hat{H}
&=&\sum\limits_{f\sigma}(\varepsilon-\mu)c^{\dagger}_{f\sigma}c_{f\sigma}
+
\sum\limits_{fm\sigma}t_{fm}c^{\dagger}_{f\sigma}c_{m\sigma}+\nonumber\\
&&+U\sum_f\hat{n}_{f\uparrow}\hat{n}_{f\downarrow}+\frac{1}{
2}\sum_{fm}V_{fm}\hat{n}_f\hat{n}_m,
\end{eqnarray}
where the last term corresponds to the allowance for energy
$V_{fm}$ of the Coulomb interaction of electrons from different
sites of the crystal lattice and $\hat{n}_f$ is the total density
operator. The last three terms in Hamiltonian
(\ref{SVHamiltonian}) together reflect the fact that the screening
radius in the systems under investigation may be by several times
larger than the unit cell parameter~\cite{Zaitsev80}. This ensures
an advantage of the Shubin-Vonsovsky model, in which the intersite
Coulomb interaction is taken into account within several
coordination spheres. In the momentum representation, Hamiltonian
(\ref{SVHamiltonian}) assumes the form
\begin{eqnarray}
\hat{H}
&=&\sum\limits_{\textbf{p}\sigma}(\varepsilon_{\textbf{p}}-\mu)
c^{\dagger}_{\textbf{p}\sigma}c_{\textbf{p}\sigma} +
U\sum_{\textbf{p}\textbf{p'}\textbf{q}}c^{\dagger}_{\textbf{p}\uparrow}
c^{\dagger}_{\textbf{p'}+\textbf{q}{\downarrow}}c_{\textbf{p}+\textbf{q}{\downarrow}}
c_{\textbf{p'}{\uparrow}}+\nonumber\\
&+&\frac12\sum_{\textbf{p}\textbf{p'}\textbf{q}\sigma\sigma'}V_{\textbf{p}-\textbf{p'}}\,c^{\dagger}_{\textbf{p}\sigma}
c^{\dagger}_{\textbf{p'}+\textbf{q}{\sigma'}}c_{\textbf{p}+\textbf{q}{\sigma'}}c_{\textbf{p'}{\sigma}},
\end{eqnarray}
where the Fourier transform of the Coulomb interaction of
electrons at the nearest sites ($V_1$) and at the next-to-nearest
sites ($V_2$) in the 2D case on the square lattice has the form
\begin{equation}\label{Vq}
V_{\textbf{q}}=2V_1(\textrm{cos}\,q_xa+\textrm{cos}\,q_ya)+4V_2
\textrm{cos}\,q_xa~\textrm{cos}\,q_ya.
\end{equation}
\begin{figure*}[t]
\begin{center}
\includegraphics[width=0.95\textwidth]{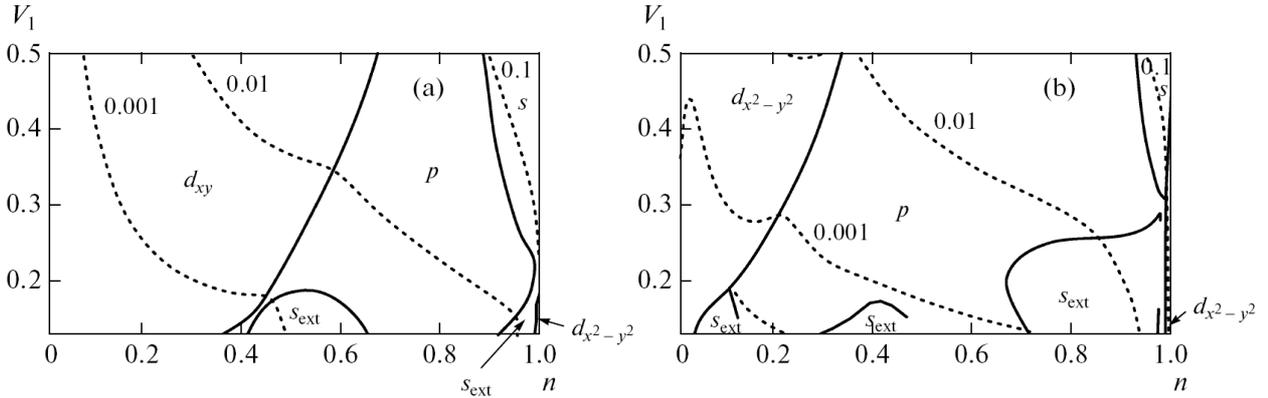}
\caption{Fig.~6. Phase diagram in the Shubin-Vonsovsky model,
constructed taking into account the second-order contributions in
$V$ for the set of parameters $t_2=t_3=0,~U=|t_1|$ and for the
ratios $V_2/V_1=0$ (a) and 0.5 (b). Dotted curves show the lines
of constant values of $\lambda$.} \label{PD2}
\end{center}
\end{figure*}

The authors of~\cite{Kagan11} contributed to the
discussion~\cite{Raghu10,Alexandrov11} by analyzing the conditions
for the occurrence of superconducting Kohn-Luttinger pairing in
the 3D and 2D Shubin-Vonsovsky models with Coulomb repulsion of
electrons at neighboring sites ($V_1\neq0,\,V_2=0$). Instead of
Yukawa potential (\ref{screening}) used as the Fourier transform
of the intersite interaction, the situation of extremely strong
Coulomb repulsion ($U\gg V_1\gg W$) was considered. In the low
electron density limit ($p_Fd\ll1$), it was shown that even in
this most unfavorable case for the occurrence of effective
attraction and superconductivity, the contribution from intersite
Coulomb repulsion $V_1$ to the effective interaction in the
$p-$wave channel is proportional to $(p_Fd)^3$ in the 3D case and
to $(p_Fd)^2$ in the 2D case in accordance with the general
quantum-mechanical results for slow particles in
vacuum~\cite{Landau89}. However, these repulsive contributions
cannot compensate contributions favoring attraction and
proportional to $(p_Fd)^2$ in the 3D case and to
$1/\ln^3[1/(p_Fd)^2]$ in the 2D case. It should be noted in this
connection that the effective attraction appears only if the
fermionic background is filled.

Thus, the previous results on Kohn-Luttinger superconducting
$p-$wave pairing being attained both in the 2D and 3D Hubbard
model with repulsion in the strong coupling limit ($U\gg W$) at
low electron density hold even when strong Coulomb repulsion of
electrons at neighboring sites ($V_1\gg W$) is included in the
Shubin-Vonsovsky model. As a result, the same expressions
(\ref{Tcp}) and (\ref{Tcp3order}) for the temperature of the
superconducting transition to the phase with $p-$wave type
symmetry, like in the absence of interstitial Coulomb repulsion
($V_1=0$), are obtained in the 3D and 2D cases. Allowance for
$V_1$ changes only the preexponential factor~\cite{Efremov00a};
therefore, superconducting $p-$wave pairing can be realized in
Fermi systems with purely Coulomb repulsion~\cite{Kagan11} in the
absence of electron-phonon interaction.

A similar analysis was carried out in~\cite{Raghu12} for the
extended Hubbard model in the Born weak coupling approximation,
and the results were the same as in~\cite{Kagan11}. Moreover, it
was noted in~\cite{Raghu12} that even in the weak coupling regime
($W>U>V$), in which controllable calculations can be performed,
the effect of long-range Coulomb interactions is suppressed in
view of the deterioration of the conditions for the evolution of
the Cooper instability. As a matter of fact, long-range
interactions in the lattice models usually contribute only to
certain pairing channels and do not affect other channels. At the
same time, the polarization contributions described by the
diagrams in Fig.~\ref{diagrams_2order} have components in all
channels, and more than one such component usually favors
attraction. In such a situation, long-range interactions probably
either do not affect at all the main component of the effective
interaction leading to pairing, or they suppress the principal
components without influencing secondary ones.

In this connection, a phase diagram was constructed
in~\cite{Raghu12} on using the extended Hubbard model with the
Kohn-Luttinger mechanism; this diagram visually reflected the
result of competition of superconducting phases with different
types of order parameter symmetry. The effective coupling constant
was calculated using the following expression for the renormalized
scattering amplitude in the Cooper channel:
\begin{eqnarray}\label{Gamma_wave_1storder}
U_{\textrm{eff}}(\textbf{p},\textbf{q})=U+V_{\textbf{p}-\textbf{q}}+U^2\Pi(\textbf{p}+\textbf{q}),
\end{eqnarray}
where $V_{\textbf{p}-\textbf{q}}$ is the Fourier transform of the
intersite Coulomb repulsion (\ref{Vq}) and
$\Pi(\textbf{p}+\textbf{q})$ is Lindhard function
(\ref{Lindhard}). Thus, the intersite Coulomb interaction $V$
in~\cite{Raghu12} was taken into account only in the first order
of perturbation theory, and the polarization contributions were
determined only by the terms of the order $U^2$. It was
shown~\cite{Raghu12} that long-range interaction has a tendency to
suppress anomalous pairing in some channels; in spite of this the
Kohn-Luttinger superconductivity survives in the entire range of
electron concentrations $0<n<1$ and for all relations between the
model parameters.

\begin{figure*}[t]
\begin{center}
\includegraphics[width=0.75\textwidth]{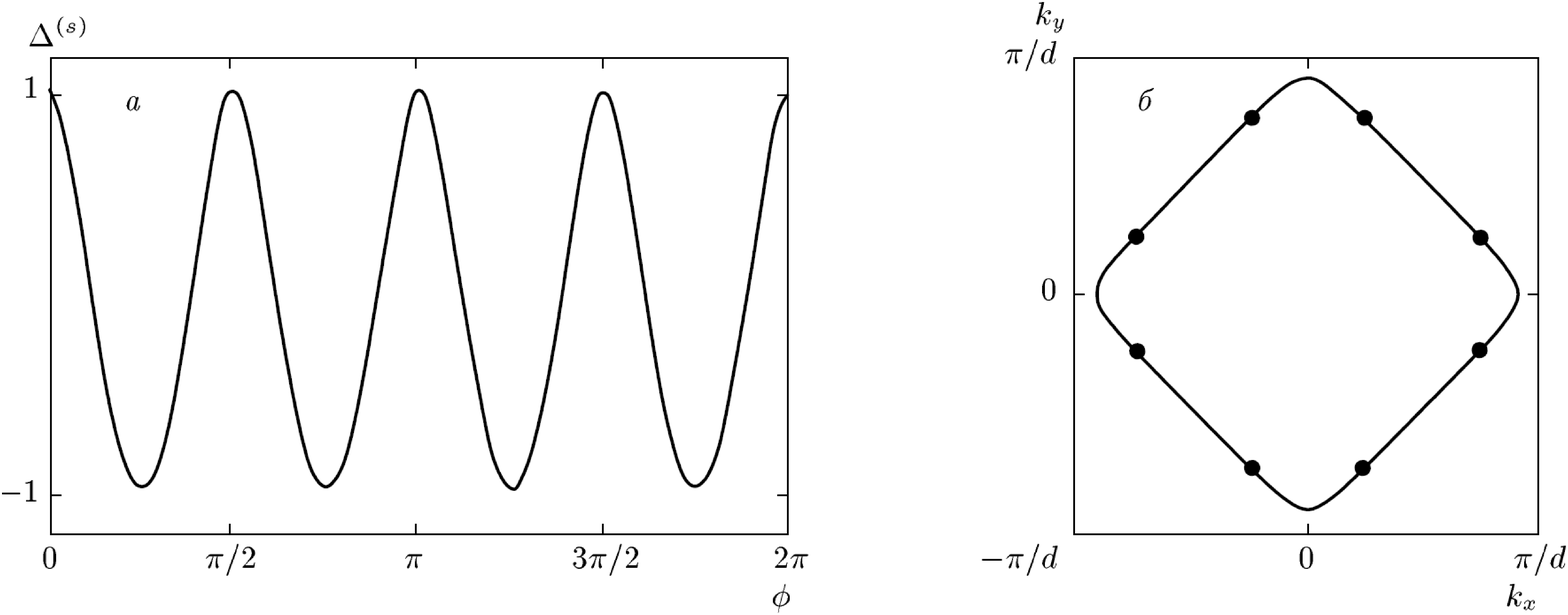}
\caption{Fig.~7. (a) Angular dependence of superconducting order
parameter $\Delta^{(s)}(\phi)$ and (b) positions of nodal points
at which $\Delta^{(s)}(\phi)$ vanishes on the Fermi contour,
calculated for parameters
$t_2=t_3=0,~U=|t_1|,~V_1=0.5|t_1|,~V_2=0,$ and $n=0.95$.}
\label{Delta_theta}
\end{center}
\end{figure*}

It was noted in~\cite{Kagan13} that effective interaction
$U_{\textrm{eff}}({\bf q})$ is characterized by a quadratic
dependence on quasi-momentum only in the range of
$\textbf{q}\textbf{d}\ll1$. Beyond this range, it is important
that the momentum dependence of $V_{\textbf{q}}$ is determined by
periodic functions. As a result, the behavior of
$U_{\textrm{eff}}({\bf q})$ is substantially modified as compared
to the momentum dependence of the Fourier transform of the Yukawa
potential. These factors considerably affect the conditions for
the Cooper instability for large electron density values, when the
Fermi surface does not exhibit spherical symmetry. Therefore, it
should be expected that the conditions for superconducting pairing
according to the Kohn-Luttinger mechanism are determined not only
by dynamic effects associated with Coulomb interactions, but also
by Brillouin zone effects.

The effect of the Coulomb interaction of electrons from the first
and second coordination spheres on the realization of the Cooper
instability was taken into account in~\cite{Kagan13} using the
Shubin-Vonsovsky model in the Born weak coupling approximation
($W>U>V$). Accordingly, in the calculating the scattering
amplitude in the Cooper channel, effective interaction
$U_{\textrm{eff}}(\textbf{p},\textbf{k})$ determined in graph form
by the sum of the five diagrams (see Fig.~\ref{diagrams_2order})
was used as the effective interaction of two electrons with
opposite values of momentum and spin. The analytic form of this
interaction is
\begin{equation}\label{Gamma_full}
U_{\textrm{eff}}(\textbf{p},\textbf{k})=U+V_{\textbf{p}-\textbf{k}}+
\delta U(\textbf{p},\textbf{k}),
\end{equation}
where the second-order corrections are given by
\begin{eqnarray}\label{Gamma_wave}
&&\delta U(\textbf{p},\textbf{k})=\frac1N\sum_{\textbf{p}_1}(U+V_{\textbf{p}-\textbf{k}})
(2V_{\textbf{p}-\textbf{k}}-V_{\textbf{p}_1+\textbf{p}}-V_{\textbf{p}_1-\textbf{k}})\nonumber\\
&&\times\frac{n_F(\varepsilon_{\textbf{p}_1})-n_F(\varepsilon_{\textbf{p}_1+\textbf{p}-\textbf{k}})}
{\varepsilon_{\textbf{p}_1}-\varepsilon_{\textbf{p}_1+\textbf{p}-\textbf{k}}}+\nonumber\\
&&+\frac1N\sum_{\textbf{p}_1}(U+V_{\textbf{p}_1-\textbf{p}})(U+V_{\textbf{p}_1-\textbf{k}})\times\nonumber\\
&&\times\frac{n_F(\varepsilon_{\textbf{p}_1})-n_F(\varepsilon_{\textbf{p}_1-\textbf{p}-\textbf{k}})}
{\varepsilon_{\textbf{p}_1-\textbf{p}-\textbf{k}}-\varepsilon_{\textbf{p}_1}}.
\label{Gamma_wave_b}
\end{eqnarray}

If the intersite Coulomb interaction is taken into account only in
the first order and only for electrons at the nearest sites
($V_1\neq0, V_2=0$ in formula (\ref{Vq})), and the excitation
spectrum is described by only one hopping parameter ($t_1\neq0,
t_2=t_3=0$), the phase diagram of superconducting states for
$U=|t_1|$ contains five regions (Fig.~\ref{PD1}). In constructing
this diagram, we used expression (\ref{Gamma_wave_1storder}) for
the effective interaction of electrons in the Cooper channel; in
this expression, the contributions proportional to $UV$ and $V^2$
and appearing in expression (\ref{Gamma_wave}) are disregarded.
The segments of the phase diagram lying on the abscissa axis
($V_1=0$) agree with the regions on the phase diagram obtained
in~\cite{Hlubina99} for the Hubbard model.

Since the first order of perturbations theory in the intersite
Coulomb interaction always has a tendency to suppress the
superconducting pairing, the possibility of the Cooper instability
realization based on the Kohn-Luttinger mechanism is associated
with the occurrence (in the second order of perturbation theory)
of contributions to the effective interaction matrix for the
Cooper channel (\ref{Gamma_wave_b}), which correspond to
attraction and are quite intense. Thus, when the Kohn-Luttinger
effects in the intersite Coulomb interaction are taken into
account, it is necessary to use complete expression
(\ref{Gamma_full}), (\ref{Gamma_wave}) for
$U_{\textrm{eff}}(\textbf{p},\textbf{q})$ and not reduced
expression (\ref{Gamma_wave_1storder}). With such an approach, the
polarization effects proportional to $UV$ and $V^2$ considerably
modify and complicate the structure of the phase diagram
(Fig.~\ref{PD2}a) even for small values of $V_1$. With increasing
parameter $V_1$ of the intersite Coulomb interaction, the value of
$|\lambda|$ for $T_c\sim W\exp(-1/|\lambda|)$ increases ($W =
8t_1$ is the 2D bandwidth for $t_2=t_3=0$). In this case, only the
three phases corresponding to the $d_{xy}-$wave, $p-$wave, and
$s-$wave types of symmetry of the superconducting order parameter
are stabilized. It should be noted that in the range of high
electron concentrations and for $0.25<V_1/|t_1|<0.5$, the
Kohn-Luttinger polarization effects lead to the occurrence of the
superconducting $s-$wave phase. This qualitative effect visually
demonstrates the importance of taking into account the
second-order processes in calculating the effective interaction of
electrons in the Cooper channel and in constructing the phase
diagram in Fig.~\ref{PD2}. Quantitative comparison of various
partial contributions to the total effective interaction shows
that $s-$wave pairing is associated with the polarization
contributions proportional to $V^2$; the main contribution in this
case for a square lattice is determined by the angular harmonic
$$g_{1}^{(s)}(\phi)=\displaystyle\frac{1}{\sqrt{\pi}}\cos 4\phi.$$
\begin{figure}[b]
\begin{center}
\includegraphics[width=0.46\textwidth]{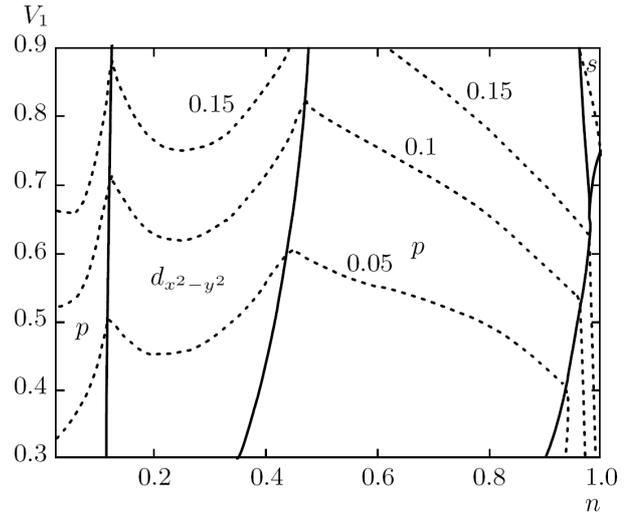}
\caption{Fig.~8. Phase diagram in the Shubin-Vonsovsky model, obtained for parameters $t_2=0.15|t_1|,~t_3=0.1|t_1|,~U=2|t_1|$, and $V_2/V_1=0.5$. Dotted curves are the lines of constant value of $\lambda$.} \label{PD5}
\end{center}
\end{figure}

Such a scenario of achieving superconducting $s-$wave pairing due
to higher angular harmonics correlates well with the experimental
data obtained recently in~\cite{Okazaki12}, in which the results
of investigating a superconductor based on iron arsenide
KFe$_2$As$_2$ using photoemission spectroscopy with ultrahigh
angular resolution were presented. It was found that this compound
is a nodal (containing gap zeros) superconductor with the $s-$wave
type symmetry of the order parameter, which has eight points at
which the gap vanishes.

Figure~\ref{Delta_theta}a shows the angular dependence of the
superconducting order parameter $\Delta^{(s)}(\phi)$,
\begin{eqnarray}
\Delta^{(s)}(\phi) &=& \frac{\Delta^{(s)}_0}{\sqrt{2}} +
\Delta^{(s)}_1\cos 4\phi + \Delta^{(s)}_2\cos 8\phi +\nonumber\\
&+&\Delta^{(s)}_3\cos 12\phi + \Delta^{(s)}_4\cos 16\phi,
\end{eqnarray}
calculated in~\cite{Kagan13} for the region of the phase diagram in which the $s-$wave pairing takes place for high electron densities. This dependence demonstrates the existence of the eight nodal points at which the gap vanishes; the position of these points on the Fermi contour (Fig.~\ref{Delta_theta}b) in calculations~\cite{Kagan13} is in qualitative agreement with the picture described in~\cite{Okazaki12}.

An analogous scenario of superconductivity realization is also observed in the $p-$wave channel; in this case, superconductivity obtained taking into account the second order of perturbation theory in the Coulomb interaction is suppressed by the initial repulsion only for the first harmonic:
\[g_0^{(p)}(\phi)=\displaystyle\frac{1}{\sqrt{\pi}}\,(A\,\textrm{sin}\,\phi+B\,\textrm{cos}\,\phi).\]
The main contribution to $\Delta^{(p)}(\hat{\textbf{p}})$  comes from the function of the next harmonic of $p-$wave pairing:
\[g_1^{(p)}(\phi)=\displaystyle\frac{1}{\sqrt{\pi}}\,(A\,\textrm{sin}\, 3\phi+B\,\textrm{cos}\,3\phi).\]

The effect of the long-range Coulomb repulsion ($V_2\neq0$) and
distant electron hoppings ($t_2\neq0,\,t_3\neq0$) on the phase
diagram of the superconducting state in the Shubin-Vonsovsky model
was also analyzed in~\cite{Kagan13}. Figure~\ref{PD5} shows the
modification of the phase diagram of the Shubin-Vonsovsky model,
which is observed upon an increase in Hubbard repulsion parameter
$U$. It can be seen that in the range of low electron densities,
as well as in the range of densities close to the Van Hove
singularity, the superconducting phase with the $d_{x^2-y^2}-$wave
symmetry of the order parameter is achieved with quite large
values of $|\lambda|\sim0.1-0.2$. This result is important for
analyzing the possibility of achieving the Kohn-Luttinger
mechanism in high-$T_c$ superconductors. It should be noted that
for $|\lambda|\sim0.2$, the superconducting transition
temperatures can reach the values $T^{d_{x^2-y^2}}_c\sim 100K$,
which are realistic for curates.

\section{$t-J$ MODEL}

\begin{figure}[t]
\begin{center}
\includegraphics[width=0.41\textwidth]{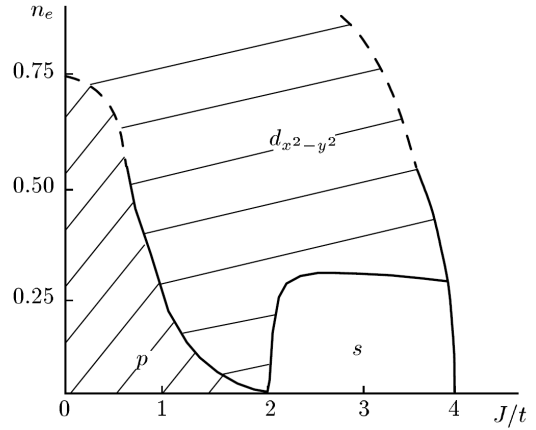}
\caption{Fig.~9. Phase diagram of the superconducting state in the 2D $t-J$ model for small and intermediate values of electron density.}
\label{tJ_PD}
\end{center}
\end{figure}

After Anderson formulated his idea \cite{Anderson87} that the
electronic properties of cuprate superconductors can be described
by the Hubbard model in the strong-coupling limit ($U\gg W$), the
so-called $t-J$ model has become extremely popular. The
Hamiltonian of the $t-J$ model with a released constraint has the
form~\cite{Kagan94_2}
\begin{eqnarray}\label{tJHamiltonian}
\hat{H}
&=&\sum\limits_{f\sigma}(\varepsilon-\mu)c^{\dagger}_{f\sigma}c_{f\sigma}
+
\sum\limits_{fm\sigma}t_{fm}c^{\dagger}_{f\sigma}c_{m\sigma}+\\
&&+U\sum_f\hat{n}_{f\uparrow}\hat{n}_{f\downarrow}+
\frac12\sum_{fm}J_{fm}\biggl(\textbf{S}_f\textbf{S}_m-\frac{\hat{n}_f\hat{n}_m}{4}\biggr).\nonumber
\end{eqnarray}
In fact, it is a model with a strong Coulomb repulsion between
electrons at the same site and with a weak antiferromagnetic
interaction $J>0$ at neighboring sites. Thus, the hierarchy of the
model parameters has the form $U\gg\{J,t\}$. The phase diagram of
the $t-J$ model constructed in~\cite{Kagan94_2} is shown in
Fig.~\ref{tJ_PD}.

For the realistic parameters of optimally doped cuprate superconductors ($J/t\sim0.5,\, n=2\varepsilon_F/W=0.85$), we can obtain the following estimate for the superconducting transition temperature:
\begin{equation}\label{Tcdx2y2_tJ}
T^{d_{x^2-y^2}}_c\sim\varepsilon_F\exp\biggl(-\frac{\pi
t}{2Jn^2}\biggr)\sim10^2\,\textrm{K}.
\end{equation}
It is important that an analogous estimate of the superconducting
transition temperature for the $d_{x^2-y^2}-$wave pairing was
obtained in~\cite{Plakida01} using a more rigorous theory for
optimally doped cuprates by employing the Hubbard operator
technique.

It should be noted that the development of the Kohn-Luttinger
ideology for the strong-coupling regime for a nearly half-filling
has become one of the most topical trends in the theory of
superconductivity in strongly correlated systems. However, the
solution of this problem requires taking into account strong
single-site correlations in all orders of perturbation theory. The
intersite correlations should be described with allowance for
second-order contributions. One of the scenarios of the
development of the theory is associated with the use of the atomic
representation~\cite{Hubbard65} and diagram technique for the
Hubbard operators~\cite{Zaitsev7576}. The models in which the
Kohn-Luttinger renormalizations can be taken into account include
the generalized $t-J-V$ model~\cite{Eremin01,Eremin12,Plakida13}
and the $t-J^*-V$ model with three-center interactions (the
important role of such interactions in describing the
superconducting state was studied
in~\cite{Hirsch89,Yushankhai90,VVV03,VVV05,Korshunov04,VVAG08,VVV08,VVV11}).
These models are effective low-energy versions of the
Shubin-Vonsovsky model.

\section{IDEALIZED MONOLAYER
OF DOPED GRAPHENE}

The popularity of the Kohn-Luttinger mechanism continues to
increase due to its possible implementation in other important
physical systems. For example, the conditions of its occurrence in
topological superfluid liquids~\cite{Marienko12}, as well as in an
idealized doped graphene monolayer, in which the effect of
nonmagnetic impurities and the Van der Waals potential of the
substrate are disregarded, are being actively discussed at
present.

One of the most interesting properties of graphene is associated
with the possibility of controlling the position of the chemical
potential in this material by applying an electric field and,
hence, by changing the type of charge carriers (electrons or
holes). It was shown experimentally~\cite{Heersche07} that short
graphene samples can be used to construct Josephson junctions by
placing them between superconducting contacts. This means that
Cooper pairs can propagate coherently in graphene. This result
suggests that graphene can probably be modified structurally or
chemically so that it becomes a magnet~\cite{Peres05} or even a
real superconductor.

It is known that, theoretically, the model with conical dispersion
requires the minimal intensity of the pairing interaction for the
development of Cooper instability~\cite{Marino06}. In this
connection, several attempts have been made to theoretically
analyze the possibility of the superconducting state in doped
graphene. The role of topological defects in Cooper pairing in
this material was studied in~\cite{Gonzalez01}. In~\cite{Uchoa07},
a phase diagram was obtained in the mean-field approximation for
the spin-singlet superconductivity in graphene; the plasmon
superconductivity mechanism leading to low superconducting
transition temperatures in the $s-$wave channel was investigated
for realistic values of electron concentrations. The possibility
of inducing superconductivity in graphene by electron correlations
was investigated in~\cite{Black07,Honerkamp08}.
In~\cite{Kiesel12}, the functional renormalization group method
was employed to study the competition between the superconducting
phase with the $d + id-$symmetry type of the order parameter and
the phase of the spin density wave on the Van Hove singularity in
the density of electron states of graphene. In the vicinity of the
Van Hove singularity, superconducting phases with the $d +
id-$wave and $f-$wave types of the order parameter symmetry were
found.

In~\cite{Gonzalez08}, the situation was considered with the Fermi
level near one of the Van Hove singularities in the density of
states of graphene. It is well known that these singularities can
enhance magnetic and superconducting
fluctuations~\cite{Markiewicz97}. According to the scenario
described in~\cite{Gonzalez08}, the Cooper instability appears due
to anisotropy of the Fermi contour for Van Hove filling $n_{VH}$,
which in fact is related to the Kohn-Luttinger mechanism. It was
noted~\cite{Gonzalez08} that the implementation of this mechanism
in graphene is possible because the electron-electron scattering
becomes strongly anisotropic and, hence, a channel with attraction
may appear for some harmonics with a non-trivial angular
dependence on the Fermi surface. Such a Cooper instability in an
idealized graphene monolayer can ensure superconducting transition
temperatures up to $T_c\sim 1\,K$ depending on the ability to tune
the chemical potential level to the Van Hove singularity to the
greatest possible extent. It should be noted that only the Coulomb
repulsion of electrons at one site was taken into account in
calculations. As mentioned above, the existence of the Van der
Waals potential of the substrate and nonmagnetic impurities were
ignored.
\begin{figure}[t]
\begin{center}
\includegraphics[width=0.45\textwidth]{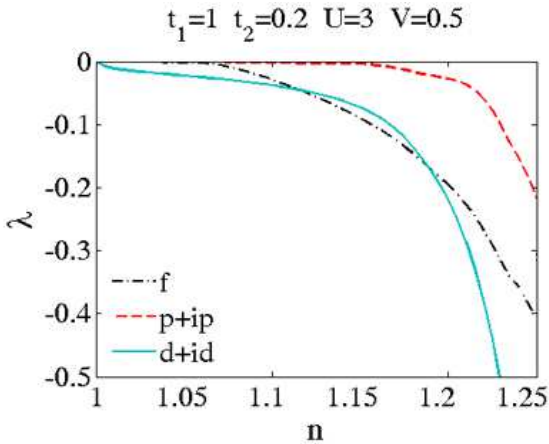}
\caption{Fig.~10. Dependence of $\lambda$ on carrier concentration
$n$ taking into account the effective interaction of electrons
with energies corresponding to both branches of the graphene
spectrum for $t_2=0.2|t_1|,\,U=3|t_1|$ and
$V=0.5|t_1|$}\label{lambdas_ab_t2}
\end{center}
\end{figure}

The possibility of competition and coexistence of the Pomeranchuk
instability and the Kohn--Luttinger superconducting instability in
graphene was considered in~\cite{Valenzuela08}.
In~\cite{Nandkishore12a,Nandkishore12b,Nandkishore12c}, it was
shown by the Kohn--Luttinger mechanism that chiral
superconductivity of the $d + id-$wave type can be achieved in a
doped graphene monolayer. Using the renormalization group method,
the authors of~\cite{Nandkishore12a,Nandkishore12b,Nandkishore12c}
in fact proved that the Cooper instability evolves simultaneously
in two degenerate $d-$wave channels.

Our recent publication~\cite{Kagan14} was devoted to analyzing
Kohn--Luttinger superconductivity in an idealized doped graphene
monolayer taking into account the Coulomb repulsion of electrons
on the same and nearest carbon atoms in the Born weak coupling
approximation. The necessity of taking into account the long-range
Coulomb interaction in calculating the physical characteristics
was dictated by the results of recent work~\cite{Wehling11}, in
which the partly screened frequency-dependent Coulomb interaction
was calculated ab initio in constructing the effective many-body
model of graphene and graphite. It was found that the one-atomic
repulsion in graphene amounts to $U = 9.3$ eV, which contradicts
the intuitively predicted small value of $U$ and weak coupling ($U
< W$). Calculations demonstrated the fundamental importance of
taking into account nonlocal Coulomb interaction in graphene
because the Coulomb repulsion of electrons located at neighboring
sites is $V = 5.5$ eV according to ab initio
calculations~\cite{Wehling11}. It should be noted that other
researchers consider the value of $V$ to be much smaller.

In the hexagonal lattice of graphene, each unit cell corresponds
to two carbon atoms; therefore, the entire lattice can be split
into two sublattices $A$ and $B$. The Hamiltonian of the
Shubin--Vonsovsky model for graphene, which takes into account
electrons hoppings between the nearest and next-to-nearest atoms,
as well as the Coulomb repulsion of electrons on the same and on
neighboring atoms, has the following form in the Wannier
representations:
\begin{eqnarray}\label{grapheneHamiltonian}
\hat{H}&=&\hat{H}_0+\hat{H}_{int},
\end{eqnarray}
where in the real space
\begin{eqnarray}
\hat{H}_0&=&-\mu\sum_{f} (\hat{n}^{A}_{f}+
\hat{n}^{B}_{f})-t_1\sum_{\langle
fm\rangle\sigma}(a^{\dag}_{f\sigma}b_{m\sigma}+\textrm{h.c.})\nonumber\\
&&-t_2\sum_{\langle\langle
fm\rangle\rangle\sigma}(a^{\dag}_{f\sigma}a_{m\sigma}+
b^{\dag}_{f,\sigma}b_{m,\sigma}+\textrm{h.c.}),\label{H0}\\
\hat{H}_{int}&=&U\sum_{f}
(\hat{n}^{A}_{f\uparrow}\hat{n}^{A}_{f\downarrow}+
\hat{n}^{B}_{f\uparrow}\hat{n}^{B}_{f\downarrow})+V\sum_{\langle
fm\rangle} \hat{n}^{A}_{f}\hat{n}^{B}_{m}.\label{Hint}
\end{eqnarray}
Operators $a^{\dag}_{f\sigma}(a_{f\sigma})$ in expressions
(\ref{H0}) and (\ref{Hint}) create (annihilate) an electron with
spin projection $\sigma=\pm1/2$ at site $f$ of sublattice $A$. At
the same time, expression
$$\displaystyle\hat{n}^{A}_{f}=\sum_{\sigma}\hat{n}^{A}_{f\sigma}=
\sum_{\sigma}a^{\dag}_{f\sigma}a_{f\sigma}$$ denotes the operator
of fermion density at site $f$ of sublattice $A$ (analogous
notation can be used for sublattice $B$). In Hamiltonian
(\ref{grapheneHamiltonian})--(\ref{Hint}), angle brackets
$\langle...\rangle$ indicate that the summation is carried out
only over the nearest neighbors, while
$\langle\langle...\rangle\rangle$ indicates that the summation is
performed over the next-to-nearest neighbors.

Passing to the momentum space and performing the Bogoliubov $u-v$
transformation,
\begin{eqnarray}\label{uv}
&&a_{\textbf{k}\sigma}=w_{\textbf{k}}\alpha_{\textbf{k}\sigma}+z_{\textbf{k}}\beta_{\textbf{k}\sigma},\nonumber\\
&&b_{\textbf{k}\sigma}=w^*_{\textbf{k}}\beta_{\textbf{k}\sigma}-z^*_{\textbf{k}}\alpha_{\textbf{k}\sigma},
\end{eqnarray}
where $\alpha_{\textbf{k}\sigma}$ and $\beta_{\textbf{k}\sigma}$
are the operators describing the dynamics of electrons in the
upper and lower bands of graphene, we can diagonalize Hamiltonian
$\hat{H}_0$. The interacting part of the Hamiltonian
$\hat{H}_{int}$ was written in~\cite{Kagan14} in the
representation of Bogoliubov operators (\ref{uv}); we derived the
expression for the effective interaction of electrons taking into
account the polarization contributions described by the diagrams
in Fig. 2.
\begin{figure}[t]
\begin{center}
\includegraphics[width=0.48\textwidth]{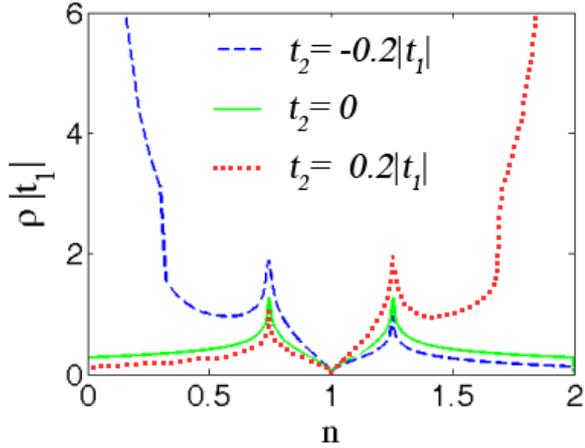}
\caption{Fig.~11. Modification of the electron density of states
in graphene with the inclusion of hoppings to the next-to-nearest
atoms for $t_2 = -0.2$ (dashed curve), $t_2 = 0$ (solid curve),
$t_2 = 0.2$ (dotted curve).}\label{DOS_graphene}
\end{center}
\end{figure}

The possibility of Cooper pairing is determined by the
characteristics of the energy spectrum in the vicinity of the
Fermi level and by the effective interaction of electrons near the
Fermi surface~\cite{Gor'kov61}. We assumed in~\cite{Kagan14} that
upon doping of graphene, the chemical potential gets into the
upper band; accordingly, in analyzing the conditions for anomalous
pairing, we considered the polarization contributions associated
with the Coulomb interaction of electrons with energies
corresponding to only one or both branches of the energy spectrum
of graphene (both Dirac cones).

Figure 10 shows the dependences of the effective coupling constant
on the electron concentration, which were obtained taking into
account the effective interaction of electrons with the energies
corresponding to both branches of the graphene energy spectrum for
the set of parameters $t_2 = 0.2,\,U = 3$, and $V = 0.5$. It can
be seen that for electron densities $1 < n < 1.13$, competition
appears between superconducting phase of the $d + id-$wave type of
the order parameter symmetry, which are described by 2D
representation $E_2$, and the superconducting phase with the
$f-$wave type of symmetry. For electron concentrations
$1.18<n<1.25$, the ground state of the system corresponds to the
superconducting phase with the $d + id-$wave type symmetry of
order parameter.

Analysis carried out in~\cite{Kagan14} revealed that the inclusion
of electron hoppings to next-to-nearest carbon atoms ($t_2$) does
not qualitatively affect the competition between superconducting
phases. Such a behavior of the system can be explained by the fact
that activation of hoppings $t_2 > 0$ or $t_2 < 0$ leads to only a
quantitative change in the electron density of states in graphene,
but does not affect its dependence on the carrier concentration
(Fig. 11). As a result, allowance for distant hoppings in $t_2$
leads to an increase in the absolute values of the effective
interaction and, hence, to a higher superconducting transition
temperature in doped graphene~\cite{Kagan14}.

The possibility of Cooper pairing in graphene was analyzed
in~\cite{Zaitsev11} in the opposite strong-coupling limit $U\gg t$
on the basis of the kinematic mechanism of superconductivity using
the diagram technique for the Hubbard
operators~\cite{Zaitsev7576,VVVSGO01}. As mentioned above, the
feasibility of the strong-coupling limit for graphene was
announced in~\cite{Wehling11}.

\section{CONCLUSIONS}

In this review, we have demonstrated the instability of the normal
state of a repulsive electron gas and of electron systems on the
lattice to the transition to the superconducting phase in
accordance with the Kohn--Luttinger mechanism for various electron
models. The initial conclusion concerning Cooper instability for
the model of a repulsive Fermi gas with a quadratic dispersion
relation was generalized for electrons in real crystalline solids
considered in the tight binding approximation. The difference
between the dispersion relation for electrons from the quadratic
law leads to a number of additional peculiarities associated with
the effects of the Brillouin zone. For example, it turned out that
the form of the electron energy spectrum determined by hopping
parameters affects the concentration dependence of the
superconducting transition temperature as well as the order
parameter symmetry. As a result, there is a change in the
structure of the phase diagram that determines the regions of
achievement of superconducting phases with different types of the
order parameter symmetry. However, the conclusion concerning the
possibility of Cooper instability with the Kohn--Luttinger
mechanism in the electron plasma in the tight binding
approximation generally holds.

In this review, it was illustrated that the universal nature of
the Kohn--Luttinger mechanism is preserved even if we take into
account the finiteness of the screening radius in repulsive Fermi
systems. At the same time, investigations based on the
Shubin--Vonsovsky model demonstrated that it is important to take
into account the Coulomb repulsion of electrons at different sites
of crystal lattice. In this case, the phase diagram of the
superconducting state changes, and the superconducting transition
temperature can be increased under certain conditions.

We also showed that the Kohn--Luttinger mechanism of
superconducting pairing can be realized in systems with a linear
dispersion relation. This was demonstrated for an idealized
graphene monolayer possessing a hexagonal lattice with two carbon
atoms per unit cell. It was shown that the polarization effects in
such a system lead to an effective attraction in the Cooper
channel.

The above arguments lead to the conclusion about the universal
nature of the Kohn--Luttinger mechanism for the formation of
Cooper instability in repulsive Fermi systems and for
superconducting pairing with a nonzero orbital angular momentum.
It should also be noted that in many cases this mechanism leads to
quite high superconducting transition temperatures (as shown in
[40], especially in the two-band situation with a wide and a
narrow band). Moreover, for electron concentrations close to the
Van Hove singularity in the electron density of states, the
superconducting transition temperatures increase still further and
may reach the values of the order of $100\,K$ even in the one-band
case for intermediate values of the ratio of the Hubbard repulsion
parameter to the conduction band width ($U/W$).

\section*{ACKNOWLEDGMENTS}

The authors are grateful to the late A.\,S. Aleksandrov, D.\,V.
Efremov, V.\,V. Kabanov, the late Yu.\,V. Kopaev, K.\,I. Kugel,
M.\,S. Mar'enko, N.\,M. Plakida, and A.\,V. Chubukov for numerous
discussions and unceasing interest in this research.

This study was supported by the Program of the Physics Department
of the Russian Academy of Sciences (project no. P.3.1) and the
Russian Foundation for Basic Research (project nos. 14-02-00058
and 14-02-31237). The work of two coauthors (M.\,M.\,K. and
V.\,A.\,M) was supported financially by grant no. MK--526.2013.2
of the President of the Russian Federation and the Dynasty
Foundation.


\begin{thebibliography}{99}
\bibitem{BCS57}
J. Bardeen, L. Cooper, and J. Schrieffer, Phys. Rev. \textbf{108},
1175 (1957).

\bibitem{Anderson87}
P.\,W. Anderson, Science {\bf 235}, 1196 (1987).

\bibitem{Vollhardt90}
D. Vollhardt and P. Woelfle, \emph{The superfluid phases of Helium
3}, Taylor and Francis, London, 1990.

\bibitem{Volovik92}
G.\,E. Volovik, \emph{Exotic properties of superfluid $^3$He},
World Scientific, Singapore, 1992.

\bibitem{Volovik03}
G.\,E. Volovik, \emph{The Universe in a Helium Droplet}, Clarendon
Press, Oxford, 2003.

\bibitem{Regal03}
C.\,A. Regal, C. Ticknor, J.\,L. Bohn, and D.\,S. Jin, Phys. Rev.
Lett. \textbf{90}, 053201 (2003).

\bibitem{Schunck05}
C.\,H. Schunck, M.\,W. Zwierlein, C.\,A. Stan, \emph{et al.},
Phys. Rev. A \textbf{71}, 045601 (2005).

\bibitem{Ott84}
H.\,R. Ott, H. Rudigier, T.\,M. Rice, \emph{et al.}, Phys. Rev.
Lett. \textbf{52}, 1915 (1984).

\bibitem{Kromer98}
S. Kromer, R. Helfrich, M. Lang, \emph{et al.}, Phys. Rev. Lett.
\textbf{81}, 4476 (1998).

\bibitem{Kuroki06}
K. Kuroki, J. Phys. Soc. Jpn. \textbf{75}, 051013 (2006).

\bibitem{Maeno01}
Y. Maeno, T.\,M. Rice, and M. Sigrist, Phys. Today \textbf{54}, 42
(2001); T.\,M. Rice and M. Sigrist, J. Phys.: Condens. Matter
\textbf{7}, L643 (1995).

\bibitem{Nagamatsu01}
J. Nagamatsu, N. Nakagawa, T. Muranaka, \emph{et al.}, Nature
\textbf{410}, 63 (2001).

\bibitem{Kamihara08}
Y. Kamihara, T. Watanabe, M. Hirano, and H. Hosono, J. Am. Chem.
Soc. \textbf{130}, 3296 (2008).


\bibitem{Novoselov04}
K.\,S. Novoselov, A.\,K. Geim, S.\,V. Morozov, \emph{et al.},
Science \textbf{306}, 666 (2004).

\bibitem{Kagan94_1}
M. Yu. Kagan, Phys. Usp. {\bf 37}, 69 (1994).

\bibitem{Lozovik08}
Yu.\,E. Lozovik, S.\,P. Merkulova, and A.\,A. Sokolik, Phys. Usp.
\textbf{51}, 727 (2008).

\bibitem{Kotov12}
V.\,N. Kotov, B. Uchoa, V.\,M. Pereira, F. Guinea, and A.\,H.
Castro Neto, Rev. Mod. Phys. \textbf{84}, 1067 (2012).

\bibitem{Wallace47}
P.\,R. Wallace, Phys. Rev. \textbf{71}, 622 (1947).

\bibitem{Castro09}
A.\,H. Castro Neto, F. Guinea, N.\,M.\,R. Peres, K.\,S. Novoselov,
and A.\,K. Geim, Rev. Mod. Phys. \textbf{81}, 109 (2009).

\bibitem{Kohn65}
W. Kohn and J.\,M. Luttinger, Phys. Rev. Lett. {\bf 15}, 524
(1965).

\bibitem{Gor'kov61}
L. P. Gor'kov and T. K. Melik-Barkhudarov, Sov. Phys. JETP {\bf
13}, 1018 (1961).

\bibitem{Freidel54}
J.\,Friedel, Adv. Phys. {\bf 3}, 446 (1954); Nuovo Cimento Suppl.
{\bf 2}, 287 (1958).

\bibitem{Lindhard54}
J. Lindhard, K. Dan. Vidensk. Selsk. Mat. Fys. Medd. \textbf{28},
8 (1954).

\bibitem{Ashcroft79}
N. Ashcroft and N. Mermin, \emph{Solid State Physics} (Holt,
Rinehart and Winston, New York, 1976; Mir, Moscow, 1979), Vol. 1.

\bibitem{Migdal58}
A. B. Migdal, Sov. Phys. JETP \textbf{7}, 996 (1958).

\bibitem{Kohn59}
W. Kohn, Phys. Rev. Lett. {\bf 2}, 393 (1959).

\bibitem{Fay68}
D. Fay and A. Layzer, Phys. Rev. Lett. {\bf 20}, 187 (1968).

\bibitem{Kagan88}
M. Yu. Kagan and A. V. Chubukov, JETP Lett. \textbf{47}, 614
(1988).

\bibitem{Baranov92b}
M.\,A. Baranov, A.\,V. Chubukov, and M.\,Yu. Kagan, Int. J. Mod.
Phys. B \textbf{6}, 2471 (1992).

\bibitem{Baranov96}
M. A. Baranov, M. Yu. Kagan, and Yu. Kagan, JETP Lett.
\textbf{64}, 301 (1996).

\bibitem{Galitskii58}
V. M. Galitskii, Sov. Phys. JETP \textbf{7}, 104 (1958).

\bibitem{Kagan89}
M. Yu. Kagan and A. V. Chubukov, JETP Lett. \textbf{50}, 517
(1989).

\bibitem{Chubukov93}
A.\,V. Chubukov, Phys. Rev. B \textbf{48}, 1097 (1993).

\bibitem{Efremov00b}
D.\,V. Efremov, M.\,S. Mar'enko, M.\,A. Baranov, and M.\,Yu.
Kagan, Physica B \textbf{284-288}, 216 (2000).

\bibitem{Bloom75}
P. Bloom, Phys. Rev. B {\bf 12}, 125 (1975).

\bibitem{Afanasiev62}
A.\,M. Afanas'ev and Yu. Kagan, Sov. Phys. JETP \textbf{16}, 1030
(1962).

\bibitem{Efremov00a}
D.\,V. Efremov, M.\,S. Mar'enko, M.\,A. Baranov, and M.\,Yu.
Kagan, JETP \textbf{90}, 861 (2000).

\bibitem{Oh94}
G.-H. Oh, Y. Ishimoto, T. Kawae, \emph{et al.}, J. Low Temp. Phys.
\textbf{95}, 525 (1994).

\bibitem{Kagan91}
M.\,Yu. Kagan, Phys. Lett. A \textbf{152}, 303 (1991).

\bibitem{KaganValkov11}
M.\,Yu. Kagan and V.\,V. Val'kov, ÆÝÒÔ {\bf 140}, 179 (2011); Low
Temp. Phys. {\bf 37}, 84 (2011); \emph{A Lifetime in Magnetism and
Superconductivity: A Tribute to Professor David Schoenberg},
Cambridge Scientific Publishers, Cambridge, 2011.

\bibitem{Woelfle11}
M.\,Yu. Kagan, V.\,V. Val'kov, and P. Woelfle, J. Low Temp. Phys.
\textbf{37}, 1046 (2011).

\bibitem{Baranov93}
M. A. Baranov, M. Yu. Kagan, and M. S. Mar'enko, JETP Lett.
\textbf{58}, 709 (1993).

%

\bibitem{Bednorz86}
J.\,G. Bednorz and K.\,A. M\"{u}ller, Z. Phys. B \textbf{64}, 189
(1986).

\bibitem{Hubbard63}
J.\,C. Hubbard, Proc. R. Soc. London A {\bf 276}, 238 (1963).

\bibitem{Izyumov94}
Yu.\,A. Izyumov, M.\,I. Katsnel'son, and Yu.\,N. Skryabin,
\emph{Magnetism of Itinerant Electrons} (Nauka, Moscow, 1994) [in
Russian].

\bibitem{Izyumov95}
Yu. A. Izyumov, Phys. Usp. \textbf{38}, 385 (1995).

\bibitem{Georges96}
A. Georges, G. Kotliar, W. Krauth, and M.\,J. Rozenberg, Rev. Mod.
Phys. \textbf{68}, 13 (1996).

\bibitem{Tasaki98}
H. Tasaki, J. Phys.: Condens. Matter. \textbf{68}, 4353 (1998).

\bibitem{VVVSGO01}
S. G. Ovchinnikov and V. V. Val'kov, \emph{Hubbard Operators in
the Theory of Strongly Correlated Electrons} (Imperial College
Press, London, 2004).

\bibitem{Baranov92a}
M.\,A. Baranov and M.\,Yu. Kagan, Z. Phys. B \textbf{86}, 237
(1992).

\bibitem{Landau89}
L. D. Landau and E. M. Lifshitz, \emph{Course of Theoretical
Physics, Volume 3: Quantum Mechanics: Non-Relativistic Theory
}(Nauka, Moscow, 1989; Butterworth– Heinemann, Oxford, 1991).

\bibitem{Scalapino86}
D.\,J. Scalapino, E. Loh, Jr., and J.\,E. Hirsch, Phys. Rev. B
\textbf{34}, 8190 (1986); \textbf{35}, 6694 (1987).

\bibitem{Kozlov89}
A. N. Kozlov, Sverkhprovodimost: Fiz., Khim., Tekh. \textbf{2}, 64
(1989).

\bibitem{Hlubina99}
R. Hlubina, Phys. Rev. B \textbf{59}, 9600 (1999); J. Mr\'{a}z and
R. Hlubina, Phys. Rev. B \textbf{67}, 174518 (2003).

\bibitem{Zanchi96}
D. Zanchi and H.\,J. Schulz, Phys. Rev. B \textbf{54}, 9509
(1996).

\bibitem{Dzyaloshinskii88}
I. E. Dzyaloshinskii and V. M. Yakovenko, Sov. Phys. JETP
\textbf{67}, 844 (1988); I. E. Dzyaloshinskii, I. M. Krichever,
and J. Chronek, Sov. Phys. JETP \textbf{67}, 1492 (1988).

\bibitem{Zheleznyak97}
A.\,T. Zheleznyak, V.\,M. Yakovenko, and I.\,E. Dzyaloshinskii,
Phys. Rev. B {\bf 55}, 3200 (1997).

\bibitem{Raghu10}
S. Raghu, S.\,A. Kivelson, and D.\,J. Scalapino, Phys. Rev. B {\bf
81}, 224505 (2010).

\bibitem{Alexandrov11}
A.\,S. Alexandrov and V.\,V. Kabanov, Phys. Rev. Lett. {\bf 106},
136403 (2011).

\bibitem{Shubin34}
S. Shubin and S. Vonsowsky, Proc. Roy. Soc. A {\bf 145}, 159
(1934); Phys. Z. Sowjetunion \textbf{7}, 292 (1935); Phys. Z.
Sowjetunion \textbf{10}, 348 (1936).

\bibitem{Vonsovsky79}
S.\,V. Vonsovsky and M.\,I. Katsnelson, J. Phys. C: Solid State
Phys. \textbf{12}, 2043 (1979); 2055 (1979).

\bibitem{Zaitsev80}
R. O. Zaitsev, Sov. Phys. JETP \textbf{51}, 671 (1980).

\bibitem{Zaitsev88}
R. O. Zaitsev, V. A. Ivanov, and Yu. V. Mikhailova, Fiz. Met.
Metalloved. \textbf{65}, 1032 (1988); R. O. Zaitsev, V. A. Ivanov,
and Yu. V. Mikhailova, Fiz. Met. Metalloved. \textbf{65}, 1108
(1989).

\bibitem{Zaitsev04}
R. O. Zaitsev, JETP \textbf{98}, 780 (2004).

\bibitem{Valkov11}
V. V. Val'kov and M. M. Korovushkin, JETP \textbf{112}, 108
(2011).

\bibitem{Kagan11}
M.\,Yu. Kagan, D.\,V. Efremov, M.\,S. Marienko, and V.\,V.
Val'kov, JETP Lett. {\bf 93}, 819 (2011).

\bibitem{Raghu12}
S. Raghu, E. Berg, A.\,V. Chubukov, and S.\,A. Kivelson, Phys.
Rev. B {\bf 85}, 024516 (2012).

\bibitem{Kagan13}
M. Yu. Kagan, V. V. Val'kov, V. A. Mitskan, and M. M. Korovushkin,
JETP Lett. \textbf{97}, 226 (2013); JETP \textbf{117}, 728 (2013).

\bibitem{Okazaki12}
K Okazaki, Y. Ota, Y. Kotani \emph{et al.}, Science {\bf 337},
1314 (2012).


\bibitem{Kagan94_2}
M.\,Yu. Kagan and T.\,M. Rice, J. Phys.: Condens. Matter {\bf 6},
3771 (1994).

\bibitem{Plakida01}
N.\,M. Plakida, JETP Lett. \textbf{74}, 36 (2001); N.\,M. Plakida,
L. Anton, S. Adam, and Gh. Adam, JETP \textbf{97}, 331 (2003).

\bibitem{Hubbard65}
J.\,C. Hubbard, Proc. R. Soc. London A {\bf 285}, 542 (1965).

\bibitem{Zaitsev7576}
R. O. Zaitsev, Sov. Phys. JETP \textbf{41}, 100 (1975);
\textbf{43}, 574 (1976).

\bibitem{Eremin01}
M. Eremin, I. Eremin, and S. Varlamov, Phys. Rev. B {\bf 64},
214512 (2001).

\bibitem{Eremin12}
M.\,V. Eremin, I.\,M. Shigapov, and I.\,M. Eremin, Eur. Phys. J. B
{\bf 85}, 131 (2012).

\bibitem{Plakida13}
N.\,M. Plakida and V.\,S. Oudovenko, Eur. Phys. J. B {\bf 86}, 115
(2013).

\bibitem{Hirsch89}
J.\,E. Hirsch, Phys. Lett. A {\bf 136}, 153 (1989).

\bibitem{Yushankhai90}
V.\,Yu. Yushankhai, G.\,M. Vujicic, and R.\,B. Zakula, Phys. Lett.
A {\bf 151}, 254 (1990).

\bibitem{VVV03}
V.\,V. Val'kov, T.\,A. Val'kova, D.\,M. Dzebisashvili, and S.\,G.
Ovchinnikov, Mod. Phys. Lett. B {\bf 17}, 441 (2003).

\bibitem{VVV05}
V. V. Val'kov and D. M. Dzebisashvili, JETP \textbf{100}, 608
(2005).

\bibitem{Korshunov04}
M. M. Korshunov, S. G. Ovchinnikov, and A. V. Sherman, JETP Lett.
\textbf{80}, 39 (2004).

\bibitem{VVAG08}
V. V. Val'kov and A. A. Golovnya, JETP \textbf{107}, 996 (2008).

\bibitem{VVV08}
V. V. Val'kov, M. M. Korovushkin, and A. F. Barabanov, JETP Lett.
\textbf{88}, 370 (2008).

\bibitem{VVV11}
V. V. Val'kov, A. A. Shklyaev, M. M. Korovushkin, and A. F.
Barabanov, Phys. Solid State \textbf{53}, 1997 (2011).

\bibitem{Marienko12}
M.\,S. Marienko, J.\,D. Sau, and S. Tewari, arXiv:1202.5784v1 26
Feb 2012.

\bibitem{Heersche07}
H.\,B. Heersche, P. Jarillo-Herrero, J.\,B. Oostinga, L.\,M.\,K.
Vandersypen, and A.\,F. Morpurgo, Nature (London) \textbf{446}, 56
(2007).

\bibitem{Peres05}
N.\,M.\,R. Peres, F. Guinea, and A. H. Castro Neto, Phys. Rev. B
\textbf{72}, 174406 (2005).

\bibitem{Marino06}
E.\,C. Marino and L.\,H.\,C.\,M. Nunes, Nucl. Phys. B
\textbf{741}, 404 (2006).

\bibitem{Gonzalez01}
J. Gonz\'{a}lez, F. Guinea, and M.\,A.\,H. Vozmediano, Phys. Rev.
B \textbf{63}, 134421 (2001).

\bibitem{Uchoa07}
B. Uchoa and A.\,H. Castro Neto, Phys. Rev. Lett. \textbf{98},
146801 (2007).

\bibitem{Black07}
A.\,M. Black-Schaffer and S. Doniach, Phys. Rev. B \textbf{75},
134512 (2007).

\bibitem{Honerkamp08}
C. Honerkamp, Phys. Rev. Lett. \textbf{100}, 146404 (2008).

\bibitem{Kiesel12}
M.\,L. Kiesel, C. Platt, W. Hanke, D.\,A. Abanin, and R. Thomale,
Phys. Rev. B \textbf{86}, 020507(R) (2012).

\bibitem{Gonzalez08}
J. Gonz\'{a}lez, Phys. Rev. B \textbf{78}, 205431 (2008).

\bibitem{Markiewicz97}
R.\,S. Markiewicz, J. Phys. Chem. Solids \textbf{58}, 1179 (1997).

\bibitem{Valenzuela08}
B. Valenzuela and M.\,A.\,H. Vozmediano, New J. Phys. \textbf{10},
113009 (2008).

\bibitem{Nandkishore12a}
R. Nandkishore, L.\,S. Levitov, and A.\,V. Chubukov, Nature Phys.
\textbf{8}, 158 (2012).

\bibitem{Nandkishore12b}
R. Nandkishore, G.-W. Chern, and A.\,V. Chubukov, Phys. Rev. Lett.
\textbf{108}, 227204 (2012).

\bibitem{Nandkishore12c}
R. Nandkishore and A.\,V. Chubukov, Phys. Rev. B \textbf{86},
115426 (2012).

\bibitem{Kagan14}
M.\,Yu. Kagan, V.\,V. Val'kov, V.\,A. Mitskan, and M.\,M.
Korovushkin, Solid State Commun. \textbf{188}, 61 (2014).

\bibitem{Wehling11}
T.\,O. Wehling, E. \c{S}a\c{s}{\i}o\u{g}lu, C. Friedrich, A.\,I.
Lichtenstein, M.\,I. Katsnelson, and S. Blugel, Phys. Rev. Lett.
\textbf{106}, 236805 (2011).

\bibitem{Zaitsev11}
R. O. Zaitsev, JETP Lett. \textbf{94}, 206 (2011); \textbf{95},
380 (2012).


\end{thebibliography}
\end{document}